\documentclass[twocolumn]{aastex631}

\newcommand\Lradio{L$_{\rm 1.4 GHz}$}
\newcommand\Lsixty{L$_{\rm 60\mu m}$}
\newcommand\Lhund{L$_{\rm 100\mu m}$}
\newcommand\zzero{$z=0$}
\newcommand\qYun{q$_{\rm Y_{01}}$}

\newcommand\noBHkappaconst{noAGN+$\kappa_{\rm const}$\,}

\newcommand\BHkappavar{AGN+$\kappa_{\rm var}$\,}
\newcommand\BHkappaconst{AGN+$\kappa_{\rm const}$\,}

\usepackage{xcolor}

\shorttitle{The FRC on FIRE: Beyond Calorimetry $\&$ Conspiracy}
\shortauthors{Ponnada et al.}
\begin{document}

\title{Hooks, Lines, and Sinkers: How AGN Feedback and Cosmic-Ray Transport shape the Far Infrared-Radio Correlation of Galaxies}

\correspondingauthor{Sam B. Ponnada}
\email{sponnada@caltech.edu}

\author[0000-0002-7484-2695]{Sam B. Ponnada}
\affiliation{TAPIR, California Institute of Technology, Mailcode 350-17, 
Pasadena, CA 91125, USA}

\author[0000-0001-8855-6107]{Rachel K. Cochrane}
\affiliation{Institute for Astronomy, University of Edinburgh, Royal Observatory, Blackford Hill, Edinburgh, EH9 3HJ, UK}
\affiliation{Department of Astronomy, 
 Columbia University, 
 New York, NY 10027, USA}

\author[0000-0003-3729-1684]{Philip F. Hopkins}
\affiliation{TAPIR, California Institute of Technology, Mailcode 350-17, 
Pasadena, CA 91125, USA}

\author[0000-0003-1257-5007]{Iryna S. Butsky}\thanks{NASA Hubble Fellow}
\affiliation{Kavli Institute for Particle Astrophysics and Cosmology and Department of Physics, 
 Stanford University, 
 Stanford, CA 94305, USA}

\author[0000-0002-3977-2724]{Sarah Wellons}
\affiliation{Department of Astronomy, Van Vleck Observatory, Wesleyan University, 96 Foss Hill Drive, Middletown, CT 06459, USA 
}

\author{N. Nicole Sanchez}
\affiliation{The Observatories of the Carnegie Institution for Science, 
813 Santa Barbara Street, Pasadena, CA 91101, USA 
}
\affiliation{TAPIR, California Institute of Technology, Mailcode 350-17, 
Pasadena, CA 91125, USA}

\author[0000-0002-3817-8133]{Cameron Hummels}
\affiliation{TAPIR, California Institute of Technology, Mailcode 350-17, 
Pasadena, CA 91125, USA}

\author{Yue Samuel Lu}
\affiliation{Department of Astronomy and Astrophysics, 
University of California 
San Diego, La Jolla, CA 92093, USA 
}

\author{Du\v{s}an Kere\v{s}}
\affiliation{Department of Astronomy and Astrophysics, 
University of California 
San Diego, La Jolla, CA 92093, USA 
}
\affiliation{Department of Physics, 
University of California 
San Diego, La Jolla, CA 92093, USA 
}

\author[0000-0003-4073-3236]{Christopher C. Hayward}
\affiliation{Center for Computational Astrophysics, Flatiron Institute, 162 Fifth Avenue, New York, NY 10010, USA}



\begin{abstract}
The far-infrared (FIR) - radio correlation (FRC) is one of the most promising empirical constraints on the role of cosmic-rays (CRs) and magnetic fields (\textbf{B}) in galaxy formation and evolution. While many theories have been proposed in order to explain the emergence and maintenance of the FRC across a gamut of galaxy properties and redshift, the non-linear physics at play remain unexplored in full complexity and cosmological context.  We present the first reproduction of the $z \sim 0$ FRC using detailed synthetic observations of state-of-the-art cosmological zoom-in simulations from the FIRE-3 suite with explicitly-evolved CR proton and electron (CRe) spectra, for three models for CR transport and multi-channel AGN feedback. In doing so, we generally verify the predictions of `calorimeter' theories at high FIR luminosities (\Lsixty\, $\gtrsim$ 10$^{9.5}$) and at low FIR luminosities (\Lsixty\, $\lesssim$ 10$^{9.5}$) the so-called `conspiracy' of increasing ultraviolet radiation escape in tandem with increasing CRe escape, and find that the global FRC is insensitive to \textit{orders-of-magnitude} locally-variable CR transport coefficients. Importantly, the indirect effect of AGN feedback on emergent observables highlights novel interpretations of outliers in the FRC. In particular, we find that in many cases, `radio-excess' objects can be better understood as \textit{IR-dim} objects with longer-lived radio contributions at low $z$ from Type Ia SNe and intermittent black hole accretion in quenching galaxies, though this is sensitive to the interplay of CR transport and AGN feedback physics. This creates characteristic evolutionary tracks leading to the $z=0$ FRC, which shape the subsequent late-time behavior of each model.

\end{abstract}



\section{Introduction}\label{sec:intro}

For over five decades, the origins of the correlation between the observed far-infrared (FIR) and radio luminosities of galaxies and its evolution with redshift has been explored with great theoretical and observational interest. First found for the cores of a few bright Seyfert galaxies \citep{van_der_kruit_observations_1971}, the FIR-radio correlation (hereafter, FRC), was established to hold for larger samples of galaxies \citep{de_jong_radio_1985,helou_thermal_1985,wunderlich_further_1987,condon_correlations_1991,yun_radio_2001,bell_estimating_2003} encompassing a broad swath of galaxy types including dwarf, irregular, and star-forming spiral galaxies. 

\citep{helou_thermal_1985} observed that the nonthermal radio emission in these systems was quickly determined to be too large to arise solely from un-resolved supernova remnants, indicating that the bulk of the emission must be arising from extended distributions of cosmic rays (CRs). Since the FIR emission originates from dust heated by star formation and AGN, the FRC presented the first sign of the close coupling between dust heating, CR production, and magnetic fields (\textbf{B}) with star formation across vastly differing galaxy conditions. 

To explain this coupling between star-formation and thermal dust emission and the non-thermal physics of CRs and \textbf{B}, \citet{voelk_correlation_1989} proposed that galaxies are simply cosmic ray electron (CRe) and FIR ``calorimeters." This meant that CRe radiate away their energy to synchrotron and Inverse Compton (IC) losses, and all of the UV photons are reprocessed in surrounding optically-thick gas to FIR. This simplified calorimeter theory \citep[and other similar theories, e.g.][]{lisenfeld_quantitative_1996} would neatly explain the FRC, however challenges naturally arise as not all galaxies are UV calorimeters \citep{bell_estimating_2003}. Indeed, the fraction of star formation that is obscured decreases with decreasing galaxy mass \citep{Hayward2014,whitaker_constant_2017}. Nor are all galaxies expected to be CRe calorimeters as the diffusive escape timescales for CRs can be much shorter than the synchrotron loss timescales in dwarf galaxies, and not strictly \textit{synchrotron} calorimeters in galaxies like the Milky Way \citep{Strong2010}, though great uncertainties remain in the transport of CRs through the interstellar and circumgalactic media (ISM and CGM) \citep{zweibel_microphysics_2013,hopkins_testing_2021,kempski_reconciling_2022,hopkins_standard_2022}.

Due to these issues with the ``calorimeter" class of models in explaining the existence of the FRC across the gamut of star-forming galaxies, \citet{helou_physical_1993} proposed an alternative non-calorimetric ``conspiracy" model, in which the CRe scale height varies as a power-law function of the gas density scale height, subject to assumptions of synchrotron losses being dominant. This class of model ameliorates the issue for galaxies with low gas surface densities; however, it fails to explain the FRC in the high surface density starburst regime and neglects several potentially important CRe loss processes. 

In a synthesis of these contrasting theoretical pictures, \citet{lacki_physics_2010} explored a large set of one-zone numerical models where galaxies are represented with a self-consistent set of galaxy-averaged parameters related to the injection of photons and CRs, computing the corresponding FIR and non-thermal properties. Through this work, a more detailed understanding of the FRC's origins emerged as a combination of UV, CRe and CR \textit{proton} calorimetry in starbursts, where rapid free-free and IC losses of CRe are balanced by the production of secondary CRe via charged pion ($\pi^{+}$ and $\pi^{-}$) decay and CRe escape is balanced by lower optical thickness to UV light. 

While the approach of one-zone, ``leaky-box" phenomenology presents an effective way to explore large parameter spaces, it by construction marginalizes over the highly complex and dynamic nature of the multi-phase ISM and CGM, which vary by several orders-of-magnitude in physical quantities relevant to CR(e) loss and propagation timescales; i.e., the local magnetic field properties, ionization fraction, turbulence, and gas and photon densities \citep{schmidt_rate_1959,vallee_magnetic_1995,Crutcher2010,Ponnada2022}. Moreover, these models fail to reproduce the observed spectral slopes at $\sim$GHz frequencies for star-forming galaxies, and fail to capture how the complex, non-linear dynamics of CRs and their back-reaction on gas may impact the relevant observables.

In the past decade, numerical simulations of galaxy formation have advanced significantly, capable of simulating the physics of star formation and feedback, magneto-hydrodynamics (MHD) \citep{peng_magnetohydrodynamic_2009,pakmor_magnetic_2014,Marinacci2014,Rieder2017,Butsky2017,Martin-Alvarez2018,Ntormousi2020,steinwandel_origin_2020,Wibking2021,robinson_regulating_2023} in concert with CRs \citep{booth_simulations_2013,salem_cosmic_2014,girichidis_launching_2016,Butsky2018,chan_cosmic_2019, buck_effects_2020,Werhahn2021,werhahn_cosmic_2021,werhahn_cosmic_2021b,pfrommer_simulating_2022,farcy_radiation-magnetohydrodynamics_2022,thomas_cosmic-ray-driven_2023}, and in many cases, halo growth from cosmological initial conditions \citep{Hopkins2020,hopkins_fire-3_2023,rodriguez_montero_impact_2024}. 

Leveraging some of these advances,  \citet{Werhahn2021} and \citet{pfrommer_simulating_2022} utilize a set of isolated galaxy simulations with CR-MHD to explore the physics driving the linear FRC and its associated scatter, largely finding agreement with the semi-analytic one-zone models of \citet{lacki_physics_2010}. These works advance our understanding of the emergence of the FRC by self-consistently evolving galactic magnetic fields and the dynamical interplay of CRs, thereby reducing the number of free parameters in the modeling, though with some caveats. Notably, they by construction utilize a steady-state formulation to solve for the relevant CRe spectra, do not explicitly resolve the multi-phase nature of the ISM, and do not evolve galaxies' cosmological environments, which in turn influence the injection of CRs from the ISM via episodic star-formation and the ensuing transport into halos with realistic, extended gas distributions. 

Evolving CR(e) spectra across several orders-of-magnitude in energy has recently become possible in galaxy formation simulations \citep{girichidis_spectrally_2020,hopkins_consistent_2022} including in cosmological simulations with explicitly resolved CR-MHD \citep{hopkins_first_2022,hopkins_fire-3_2023}. These self-consistent simulations are capable of capturing the full dynamical, non-linear, and non-equilibrium physics of CRs which may be crucial for determining the details of the intensity and spatial distributions of synchrotron-emitting gas \citep{ponnada_synchrotron_2024}.

Furthermore, the role of AGN feedback (even via strictly \textit{indirect} effects) in shaping the emergence of the FRC has remained largely unexplored in the literature. As the physics of black hole accretion and subsequent energy injection is far below modern simulation resolution limits, there are varied ``sub-grid'' prescriptions for AGN feedback in different cosmological simulation suites \citep{schaye_eagle_2015,pillepich_simulating_2018}, with the corresponding physical interpretation of the resolved-scale observables, which are typically $\gtrsim$ kpc, remaining unclear. 

Advances of late have pushed this `sub-grid' boundary for AGN energy and momentum injection much further inwards, down to \textit{$\sim$pc} scales, \textit{explicitly} evolving the known (but largely unconstrained) channels of feedback from AGN (radiative, mechanical, and non-thermal, relativistic jets) and their subsequent physical interactions with the multi-phase ISM/CGM and stellar feedback effects upon injection from a sub-grid accretion kernel scale \citep{su_which_2021,su_unraveling_2023,wellons_exploring_2023,hopkins_fire-3_2023}, with BH fueling physics therein motivated by more idealized, but hyper-refined simulations down to far smaller scales \citep{angles-alcazar_cosmological_2021}. 

The goals of this paper are to model in full cosmological complexity the emergence of the FRC across a large dynamic range in galaxy properties and characterize its evolution with redshift to the well-studied $z=0$ FRC. For the first time, we also aim to gain a better physical understanding of outliers in the FRC in the context of AGN feedback and varied CR transport. To this end, we utilize a large sample of novel zoom-in, dynamical CR-MHD simulations from the latest version of the Feedback in Realistic Environments project (FIRE)\footnote{\url{https://fire.northwestern.edu/}} simulation suite \citep{hopkins_fire-3_2023} which crucially evolve CR and CRe dynamics within a self-consistent cosmological framework while resolving multi-phase ISM/CGM structure. 

We describe our simulations, present our sample selection, and detail our post-processing methodology to generate radio continuum and multi-band UV-IR imaging in Section \ref{sec:methods}. In Section \ref{sec:FRC_on_FIRE}, we present our results and compare to the relevant observations of the FRC, and explore the origins of the FRC and its scatter in different physical regimes. With the breadth of physics probed by our simulation sample, we describe potential physical mechanisms giving rise to the  $z \sim$ 0 FRC's properties as galaxies evolve with redshift, and the emergence of interesting outliers. Finally, we discuss our results and conclusions and summarize our findings in Section \ref{sec:discussion}.

\section{Methodology}\label{sec:methods}

\subsection{Simulations}\label{sec:sims}

Our sample of FIRE-3 simulations contains galaxies with $z=0$ dark matter halo masses  M$_{\rm halo}^{z=0}$ = 3 $\times$ 10$^{\rm 10}$ - 10$^{\rm 13}$ M$_{\rm \odot}$ which were run with physics variations explored herein. In previous works utilizing FIRE simulations, these halo masses are commonly referred to as \texttt{m11}, \texttt{m12}, and \texttt{m13}, and we use the same naming convention here. These halo mass grouping correspond to M$_{\rm halo}^{z=0} \sim $ 3 $\times$ 10$^{10} - $ 7 $\times$ 10$^{11}$ M$_{\odot}$, 7 $\times$ 10$^{11}$ - 1.5 $\times$ 10$^{12}$ M$_{\odot}$, and 5 $\times$ 10$^{12}$ - 10$^{13}$ M$_{\odot}$ respectively. The baryonic mass resolution for the \texttt{m11's} ranges from m$_{\rm b}$ = 2000-10000 M$_{\rm \odot}$ depending on the individual  M$_{\rm halo}$ at $z=0$, while is fixed at 6 $\times\, 10^4$ M$_{\rm \odot}$ for \texttt{m12's} and 3 $\times\, 10^5$ M$_{\rm \odot}$ for \texttt{m13's} respectively.

These are all fully-dynamical, cosmological zoom-in, magnetohydrodynamic galaxy formation simulations capable of evolving radiation and feedback from star formation and evolution, in addition to detailed thermo-chemical properties of gas with ``live" CR spectra. These simulations are run with the \texttt{GIZMO}\footnote{GIZMO is publicly available at \url{http://www.tapir.caltech.edu/~phopkins/Site/GIZMO.html}.} code, in the mesh-free finite-mass mode. All simulations include MHD as treated in \citet{hopkins_accurate_2016,hopkins_constrained-gradient_2016}, and fully-anisotropic Spitzer-Braginskii conduction and viscosity \citep{hopkins_anisotropic_2017,Su2017}. 

Galaxy formation from cosmological initial conditions at redshifts $z\gtrsim100$ including both dark matter and baryons (in gas and stars) occurs self-consistently, with magnetic fields amplified from arbitrarily small trace seed fields at $z \approx 100$, and phase structure and thermo-chemistry in galaxies naturally emerging from cooling with temperatures $T \sim 1-10^{10}\,$K and self-gravity. Star formation occurs in self-gravitating, Jeans unstable gas with converging flows (or diverging slowly compared to the free-fall/star-formation timescale) on resolved scales. Those stars then influence the medium in turn via their injection of radiation fully-coupled to multi-band (EUV/FUV/NUV/OIR/FIR) radiation transport, stellar mass-loss, and both Type Ia and core-collapse SNe explosions (determined consonantly with up-to-date, standard stellar evolution models).

Major updates in these simulations from the FIRE-2 code version \citep{Hopkins2018} include updated explicit cooling functions and stellar evolution tracks, resulting in a better-resolved cold ISM phase, as well as a ``velocity-aware," more conservative coupling of the terminal SNe ejecta momentum (representing unresolved work in an unresolved phase of SNe evolution) to surrounding gas \citep[see][for details and discussion]{hopkins_fire-3_2023,hopkins_importance_2024}. The primary change of interest here resulting from this feedback coupling is that galaxies with halo mass (M$_{\rm halo}) \gtrsim$ 10$^{11}$M$_\odot$ have lower stellar masses compared to FIRE-2. We do not anticipate this to significantly affect the results presented here, tending to move galaxies \textit{along} the FRC at a given M$_{\ast}$, but we caution that future work referring to FIRE-3 may incorporate a different terminal SNe momentum coupling.\footnote{Which, our preliminary results show, could affect galaxy stellar mass, producing massive galaxy stellar masses similar to that of FIRE-2.}

Our sample contains cosmological realizations of 22 unique halos  whose halo, morphological, and merger histories are outlined in \citet{Hopkins2018,Hopkins2020,wellons_exploring_2023,byrne_formation_2023}. These are analyzed for three different physics variations delineated in Table \ref{tab:sims}, and we discuss the relevant physics parameters varied and their implementations below. 

We stress that our sample is not selected to be cosmologically representative, and so it is entirely possible that our results may under-/over-predict the true intrinsic scatter compared to the ``true" FRC owing to under-/over-sampling regions of cosmological parameter space in halo merger/growth histories and larger-scale (L $ \geq $ 10 Mpc, beyond the high-resolution Lagrangian volume) cosmological environments. However, our results robustly illustrate the effect of differing models for CR and AGN feedback physics in paths leading to and shaping the $z=0$ FRC, which is the scope of this work.

\begin{table}
    \centering
    \begin{tabular}{lccc}
    \hline
    Model & AGN & $\epsilon^{BH}_{\rm CR}$  & $\nu_{\rm CR}$ [s$^{-1}$] \\
    \hline
    \noBHkappaconst & $\times$ & $\times$ & 10$^{-9}$ $\beta_{\rm cr}$ R$_{GV}$$^{-0.6}$ \\ 
    \BHkappaconst & $\checkmark$ & 3$\times$10$^{-4}$ & 10$^{-9}$ $\beta_{\rm cr}$ R$_{GV}$$^{-0.6}$\\
    \BHkappavar & $\checkmark$ & 1$\times$10$^{-3}$ & $\propto$ S$_{\pm}$/$\Gamma_{\pm}$ $\sim v_{A}^{3}/\Gamma_{\pm}\ell_{A}$\\
    \hline
    \end{tabular}
    \caption{\textbf{Model properties} for the FIRE-3 simulations analyzed in this work, specifically the BH CR injection efficiency and the treatment of CR scattering. Two models include AGN feedback (with mass loading $\dot{M}_{\rm wind} = \dot{M}_{\rm BH}$, kinetic wind velocity v$_{\rm wind}$ = 3000 km s$^{-1}$, kinetic energy loading 5 $\times$ 10$^{-5}$ $\dot{M}_{BH}c^{2}$, radiative efficiency $\epsilon^{\rm BH}_{\rm r}$=0.1), but with two different models for the CR scattering rate (which determines the emergent `streaming' or `diffusion' speeds. One model does not include AGN feedback and uses the simple power-law scaling for the CR scattering rate.} 
    \label{tab:sims}
\end{table}

\subsubsection{Cosmic Ray Physics}
In the simulations used in this study, we follow CR protons and electrons from MeV to TeV energies. The CR physics is directly coupled to the dynamics, with CRs propagating along magnetic field lines according to the fully general CR transport equations \citep[see][for details of the methodology]{hopkins_consistent_2022,hopkins_first_2022}, and self-consistently includes adiabatic/convective/turbulent terms, diffusive re-acceleration, streaming/gyro-resonant loss, Coulomb, ionization, hadronic and other collisional, radioactive decay, annihilation, Bremstrahhlung, inverse Compton (IC), and synchrotron loss terms. Hadronic losses for CR protons are assumed to be dominated by the proton-proton interaction, with total pion loss rates following \citet{Mannheim1994,Guo2008} and those of \citet{Evoli2017} for antimatter. 

CRs are injected with a power-law spectrum in momentum at SNe (Types Ia \& II) and stellar winds (OB/WR) to neighboring gas cells with fixed fractions $\epsilon^{\rm inj}_{\rm CR}$ = 0.1 and $\epsilon^{\rm inj}_{\rm e}$ = 0.002 of the initial ejecta kinetic energy going into CRs (protons) and leptons (electrons). For the CR modeling, all quantities needed to compute the fully non-equilibrium CR dynamics and losses are captured in-code, except for the microphysical CR scattering rate as a function of rigidity $\nu_{\rm CR}$(R$_{\rm GV}$), where R$_{\rm GV}$ is the rigidity in gigavolts of a given CR bin (E$_{\rm CR}$/q), which ultimately drives (non-linearly via the dynamical equations) the effective diffusion and/or streaming speeds of the CRs. 

For $\nu_{\rm CR}$(R$_{\rm GV}$), we explore two model variations. One follows standard practice in CR-MHD galaxy simulations and assumes a spatially and temporally constant scaling for the scattering rate as a function of CR rigidity $\nu_{\rm CR} \sim$ 10$^{-9 }\beta_{\rm cr}$ (R$_{\rm GV}$)$^{-0.6}$ s$^{-1}$ where $\beta_{\rm cr} = v_{\rm cr}/c$ as in \citep{hopkins_first_2022}, calibrated explicitly therein to fit all of the observations of CRs in Milky-Way Solar-Circle like conditions from observations such as Voyager, AMS-02, and Fermi. This corresponds roughly in the diffusive limits to an anisotropic/parallel diffusivity $\kappa_{\rm \|} \sim {v_{\rm cr}^{2}}/{3\nu_{\rm CR}} \sim  3\times10^{29}\,\beta_{\rm cr}\,R_{\rm GV}^{0.6}$ cm$^{2}$ s$^{-1}$ (so e.g. the effective isotropic diffusivity or streaming speed of $\sim\rm{GeV}$ CR(e)s is $D_{xx} \sim 7\times 10^{28}\,{\rm cm^{2}\,s^{-1}}$ or $\sim 100\,{\rm km\,s^{-1}}$) with anisotropic streaming, advective transport and diffusive re-acceleration automatically included self-consistently via the full non-equilibrium CR flux and energy transport equations (as distinct from e.g. common pure-diffusion, pure-streaming, or Fokker-Planck type equations). We denote simulations which use this CR transport model hereafter with \textbf{$\kappa_{\rm const}$}.    

We also explore an alternative CR transport model which varies $\nu_{\rm CR}$ (and therefore the predicted diffusive and streaming speeds), motivated in principle by so-called ``extrinsic turbulence" theory \citep{jokipii_cosmic-ray_1966} as presented in \citet{hopkins_standard_2022} but with an additional modified driving term at gyroresonant scales in the sub-grid scattering prescription in order to produce reasonable $\nu_{\rm CR}$(R$_{\rm GV}$), resulting in observables in agreement with the constraints mentioned above for $z=0$ Milky-Way Solar-Circle like conditions. These runs are labeled as \textbf{$\kappa_{\rm var}$}. 

In our $\kappa_{\rm var}$ model, we utilize an empirically-motivated driving term\footnote{$S_{\rm \pm,\,ext} = (v_{A,\,{\rm ideal}}/0.007\,c)\,(v_{A,\,{\rm ideal}}/\ell_{A})\,(k_{\|}\,\ell_{A})^{-1/6}\,u_{\rm B}$} $S_{\pm}$, given by fitting CR observables such as different MeV-TeV primary and secondary spectra, B/C, proton-to-antiproton and positron-to-electron as well as radioactive isotope ratios. This driving term scales dimensionally with plasma properties proportional to the turbulent magnetic dissipation rate $\sim v_{A}^{3}/\ell_{A}$ in terms of the Alfven speed and turbulent Alfven scale $\ell_{A}$, akin to classic ``extrinsic turbulence'' type models for CR scattering \citep{jokipii_cosmic-ray_1966}, though the normalization and wavelength/scale-dependence is quite different from those classic models as required by the observations \citep[see discussion in][Section 5.3.3]{hopkins_standard_2022}. 

\subsubsection{Supermassive Black Hole/AGN Physics}
Many of the simulations explored in this work include black holes (BHs), with seeding, dynamics, accretion, and feedback physics described extensively in \citet{wellons_exploring_2023,hopkins_fire-3_2023}. BHs are randomly seeded from star-forming gas preferentially at high surface densities and low metallicites, and are permitted to merge if sufficiently close and gravitationally bound. Accretion from the ISM into the BH accretion reservoir is continuous with an accretion efficiency parameter calibrated from much higher-resolution simulations on smaller scales than resolved here that represent the effects of `gravitational torques' driving accretion \citep{angles-alcazar_cosmological_2021}, and flows from the accretion reservoir onto the BH on a depletion timescale motivated by a \citet{Shakura_Sunyaev1973} $\alpha$-disk. 

Feedback from BHs consistently follows radiative, mechanical (kinetic), and CRs in the form of relativistic jets coupled to the general CR-MHD solver beyond the BH accretion kernel scale. 

Radiative feedback is given by the accretion disc emitting a bolometric luminosity of

\begin{equation}
    \dot{E}^{\rm BH}_{\rm rad} \equiv L_{\rm bol} = \epsilon^{BH}_{r} \dot{M}_{\rm BH}c^{2}
\end{equation}

with $\epsilon^{\rm BH}_{\rm r}$ = 0.1 and the total photon momentum flux given by $\dot{p}$$_{rad}$ = $L_{\rm abs}/c$, where $L_{\rm abs}$ is the photon luminosity absorbed by a given gas element. This radiative feedback is injected at the BH kernel location and transported according to the same locally extincted RHD approximation as the radiative feedback from stars \citep{hopkins_radiative_2020}.

Non-relativistic mechanical feedback in the form of outflows from the accretion disc are modeled using a hyper-refinement particle-spawning scheme with particle resolution $\sim$1000 times higher (lower mass) than the typical gas cell in the simulation in order to effectively capture reverse shocks, and are de-refined when fully mixed with the ambient surrounding gas. These outflows have initial positions and velocities aligned with the spin axis of the BH, with kinetic wind velocities v$_{\rm wind}$ = 3000 km s$^{-1}$ and mass-loading $\dot{M}_{\rm wind, BH}$ = $\dot{M}_{\rm BH}$ which yield a kinetic energy loading $\dot{E}_{\rm wind}^{\rm BH} \approx  $ 5 $\times$ 10$^{-5}$ $\dot{M}_{\rm BH}$c$^{2}$.

Since all of the simulations including BHs here also consider CR feedback, we model relativistic mechanical feedback in the form of jets by injecting CRs to the hyper-refined mechanical feedback cells, with injection properties identical to stellar CRs up to the injection efficiency, which is given by 

\begin{equation}
    \dot{E}_{\rm CR}^{\rm BH} \equiv \epsilon_{\rm CR}^{\rm BH} \dot{M}_{\rm BH} c^{2}
\end{equation}

with fiducial $\epsilon_{\rm CR}^{\rm BH}$ = 3 $\times$ 10$^{-4}$ for $\kappa_{\rm const}$ runs and $\epsilon_{\rm CR}^{\rm BH}$ = 10$^{-3}$ for $\kappa_{\rm var}$ runs.

For all three feedback channels, the equivalent rest mass energy is removed from the accretion reservoir of the BH particle, without the inclusion of a BH mass limit or further accretion dependence. The choices of accretion energy conversion efficiencies are motivated by extensive parameter studies in \citet{wellons_exploring_2023} for models capable of reproducing galaxies which are reasonable on various population-wide scaling relations, as well as more detailed observational constraints \citep{byrne_formation_2023}.

\subsection{Simulation Post-Processing}
For our fiducial results, we center the galaxies on the stellar center-of-mass and orient the galaxies to be face-on with the the angular momentum of the neutral gas as the $\hat{z}$ axis.  All vector fields ($\textbf{r}$, $\textbf{B}$, $\textbf{v}$) are transformed accordingly. 

The production of the subsequent images is done after selecting all star and gas particles with positions $|x|$, $|y|$ $\leq$ 0.15 R$_{\rm 200}$ and $|z| \leq$ 200 kpc, and all images are generated at a 50 pc/pixel resolution.

We heavily emphasize that for both our radio continuum and far-infrared images, we do not include any template spectrum for the emission arising from the AGN for simulations including BHs as these vary widely with BH mass, inclination angle, and accretion rate, and would be sensitive to radiative properties on scales far below our resolution limit here. All AGN feedback modes included in the simulations are described in the above sections, and any ensuing effects on the radiative properties arise solely from the physical coupling of the radiative, kinetic, and mechanical feedback to gas. 

Thus, the influence of AGN feedback on our results is in a more \textit{indirect} manner owing to its interplay with ISM/CGM cooling and dynamics and cosmological galaxy-formation physics from the deposition scale of the BH interaction kernel ($\sim$ 10 pc from the BH itself) on the surrounding gas, which is of $\gtrsim$ pc resolution rather than through ``direct'' contributions via disk-, cocoon-, or jet-dominated radio/FIR emission.

\subsubsection{Radio Continuum Emission}\label{sec:Radio}
For the computation of the optically thin, non-thermal radio emission from our simulations, we follow the same exact procedure as in \citet{ponnada_synchrotron_2024}, where for each gas cell in our simulations, synchrotron emissivities are calculated from the internally evolved CRe spectra, $j_e(E)$, and components of the magnetic field perpendicular to the line of sight for a given galactic orientation, $\textbf{B}_{\bot}$. 

Then, we compute the critical frequency of emission for each spectral bin of CRe,    
        \begin{equation}
        \nu_c(B_{\bot},E) = \frac{3eB_{\bot}}{{4 \pi m_{e}c}} \left(\frac{E}{m_{e}c^{2}}\right)^2
        \end{equation}

where m$_{\rm e}$ is the electron mass, and c is the speed of light. We then compute the specific emissivities of each gas cell by integrating over j$_{e}$ to determine the contribution from each energy bin at a given frequency, and finally produce Stokes I by integrating these specific emissivities along the line of sight using a projection routine first described in \citet{Hopkins2005}. Synchrotron self-absorption is not modeled as it is unimportant for radio continuum emission from galaxies at low brightness temperatures, though may be important for AGN jet template spectra not included here \citep{condon_radio_1992}.

\subsubsection{Far-Infrared Images}\label{sec:SKIRT}
The production of the far-infrared images follows the procedure of \citet{cochrane_predictions_2019,Cochrane2023a,cochrane_impact_2023}, using the radiative transfer code \texttt{SKIRT}\footnote{\url{http://www.skirt.ugent.be}} Version 8 \citep{baes_efficient_2011,camps_skirt_2015}. Predictions for the rest-frame ultraviolet (UV) to far-infrared (FIR) wavelengths are generated from the particle cut specified above. We assume a constant dust-to-metals ratio (D/Z) of 0.4 \citep{Dwek1998}, with metallicity evolved for each gas cell self-consistently, and dust destruction at T $>$ 10$^{6}$ K \citep{Draine1979} for a graphite, silicate, and PAH dust mixture following the \citet{Weingartner2001} Milky Way dust model. We caution that the assumption of constant D/Z  may not hold for dwarfs \citep{2014A&A...563A..31R,Choban2024}, which may move them to slightly lower FIR luminosities. The SEDs and IMFs of \citet{BruzualCharlot2003} are adopted for star particles, depending on their ages and metallicities.

The radiative transfer calculation is done using 10$^{6}$ photon packets on an octree dust grid where cell sizes are adjusted by the dust density distribution, constrained such that no dust cell may contain great than 10$^{-4}$ per cent of the galaxy's total dust mass, and outputs global galaxy spectral energy distributions (SEDs) and images at the same fiducial pixel resolution. For extensive convergence tests of these parameter choices for our simulations, we refer the reader to the appendices in \citet{Cochrane2023a}.

\subsection{Synthetic-Observational Sample Selection}\label{sec:sample}

We select all snapshots with \Lsixty\, $\geq$ 10$^{7.5}$ L$_{\odot}$ and \Lradio\, $\geq$ 10$^{18.5}$ W \,m$^{-2}$ as a synthetic-observable parameter space to correspond roughly to the luminosity limits of observational studies of the $z=0$ FRC. As a result, the majority of of our mock-observed ``detections'' draws from snapshots of relatively more massive halos (M$_{\rm halo}^{z=0} \gtrsim$ 5 $\times$ 10$^{11}$ M$_{\odot}$), though lower mass halos populate the mock-observed space at low-\Lsixty ($\lesssim$ 10$^{9.5}$ L$_{\odot}$). Since some of our simulation suite has snapshots sampled at higher time resolution (typically for the higher-resolution, m$_{\rm b}$ = 2 $\times$ 10$^{3}$ - 10$^{4}$ M$_{\odot}$, \texttt{m11's} classical to intermediate dwarf galaxy simulations), we down-sample those to the same time sampling as the more sparsely snapshot sampled simulations to avoid biasing our conclusions. In total, we analyze over 8750 simulation snapshots for this work, with $\sim$3100 snapshots meeting the mock-observational selection criteria, which is reduced to 1375 after the down-sampling of high time-resolution runs.

Finally, we invoke an additional cut in order to exclude snapshots that might plausibly be overly-contaminated by AGN accretion activity for our runs including BHs: we remove those which have $\frac{L^{\rm AGN}_{\rm bol}}{30}$ $>$ \Lsixty$^{\rm stars}$, where $L_{\rm AGN}^{\rm bol} \equiv 0.1\,\dot{M_{\rm BH}} c^{2}$ is the AGN bolometric luminosity used in-code, and \Lsixty$^{\rm stars}$ is the stellar luminosity at $60\,{\rm \mu m}$ predicted by our radiative transfer post-processing without including any AGN. The factor of $30$ accounts for the typical bolometric correction for typical type I (broad-line) quasars at this same wavelength \citep{richards_spectral_2006}. 

In other words, we remove any snapshot where we would expect the AGN to potentially dominate the FIR luminosities we will study, as they would be removed in observational samples. This amounts to just 53 snapshots in our synthetically-observed sample ($\sim$3.5\% of the total), and this selection cleans our sample from AGN contamination in the regime where AGN are expected to dominate the FIR as well as the radio continuum, given the observed normalization of the FRC for AGN \citep{van_der_kruit_observations_1971,de_jong_observations_1971,delhaize_vla-cosmos_2017}. Our results are not particularly sensitive to this cut at all, let alone to the precise ratio of luminosities used for the cut above. 

\begin{figure}
    \centering
    \includegraphics[width=0.5\textwidth]{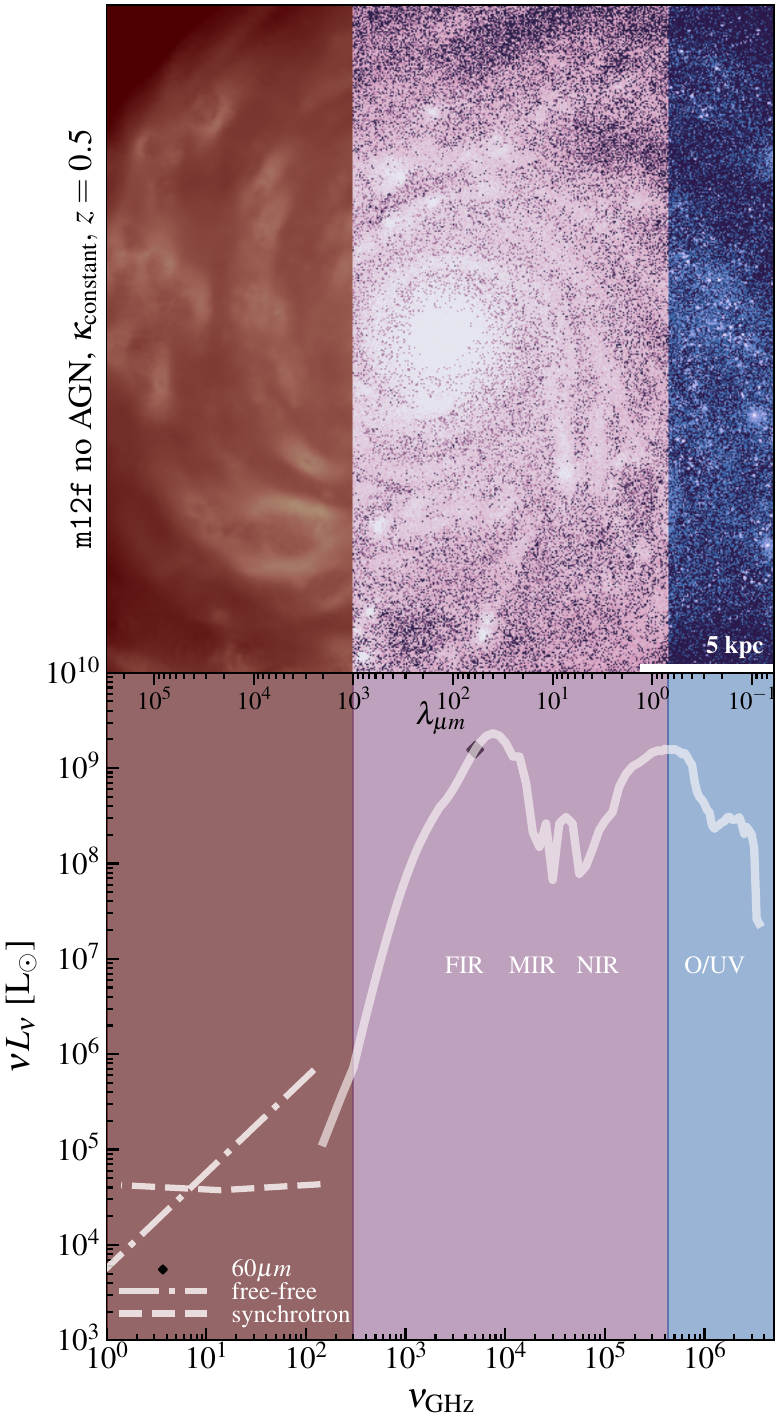}
    \caption{\textit{Synthetic observations from radio to UV:} \textbf{Top}: radio continuum image at 140 MHz (\textbf{left}) generated from the procedure in Section \ref{sec:Radio}, FIR image at 60$\mu m$ (\textbf{center}), and FUV image at 150 nm (\textbf{right}), both generated using \texttt{SKIRT} \citep{camps_skirt_2015} with procedure described in Section \ref{sec:SKIRT} for \texttt{m12f}, a Milky-Way-like simulated galaxy at $z=0.5$. \textbf{Bottom}: Corresponding SED ($\nu L_{\rm \nu}$) in L$_{\rm \odot}$ from 1.4 GHz to 85 nm, with optical/UV to FIR  (solid), thermal free-free radio  (dot-dashed), and non-thermal synchrotron (dashed) emission computed directly from self-consistently evolved magnetized gas, stellar, and CR properties. Our synthetic SEDs bear remarkable similarity to those observed across over six orders-of-magnitude in frequency/wavelength space.}
    \label{fig:sed_demo}
\end{figure}

In Figure \ref{fig:sed_demo}, we show an illustrative example of the multi-wavelength image generation done for all of the simulated galaxies in our sample from radio to optical/UV frequencies. The synthetically-observed SEDs are similar in shape and normalization to those observed \citep{Smith2012,Tabatabaei2017}.

\section{The Far-Infrared Radio Correlation on FIRE}\label{sec:FRC_on_FIRE}
\begin{figure*}
    \centering
    \includegraphics[width=1.0\textwidth]{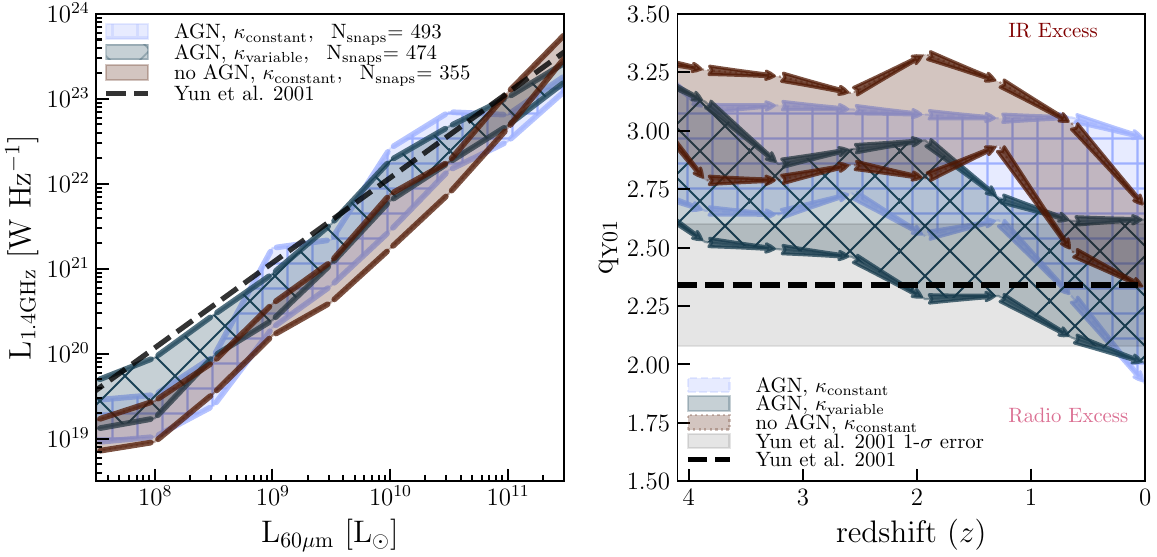}
    \caption{\textit{The FRC for FIRE-3 simulations from $z$=0-5.} \textbf{Left:} \Lradio  vs.  \Lsixty for all snapshots analyzed in this study, with \noBHkappaconst, \BHkappaconst, \BHkappavar in brown with no hatching, light blue with square hatching, and navy with diamond hatching respectively, with the observed $z=0$ relation of \citep{yun_radio_2001} for star-forming galaxies shown as the black dashed line. Hatched and shaded regions show the 32-68 percentile (approximate $\sim 1\sigma$) confidence intervals for equally spaced bins in  \Lsixty, with line segments demarcating bins. Regardless of physics variation (AGN vs. no AGN, $\kappa_{\rm const}$ vs. $\kappa_{\rm var}$), our simulations generally match the observed $z$=0 correlation at high \Lsixty (\Lsixty $\geq$ 10$^{9.5}$ L$_{\odot}$), with a weakly super-linear relation at low \Lsixty (\Lsixty $\leq$ 10$^{9.5}$ L$_{\odot}$) arising from a transition to super-linear L$_{\rm IR}$-SFR and  L$_{\rm 1.4\, GHz}$-SFR correlations. \textbf{Right:} \qYun vs. redshift for the same snapshots as the left, here shown in equally-spaced redshift bins, with arrow segments demarcating bins. While all physics variations reach the observed $z=0$ value of \qYun on average, the evolution of \qYun and its scatter varies with redshift owing to the interplay of AGN and CR transport physics, which is broken down further in Figure \ref{fig:q_z_mass}.} 
    \label{fig:FRC_all}
\end{figure*}

The FRC from $z=5$ to $z=0$ for the synthetic observations described in Section \ref{sec:sample} is shown in Figure \ref{fig:FRC_all}. Since there is qualitatively little evolution in the FRC across this redshift range for the sample selected, we plot all the synthetic observations together in the left panel, and show the redshift evolution of the FIR-radio flux ratios in the right panel.

Here, we compare the equivalent synthetic-observational quantities to those of \citet{yun_radio_2001} (hereafter Y01). Namely, we utilize the definition of the flux ratio of the 60-100$\mu m$ FIR emission to 1.4 GHz radio emission, hereafter denoted as \qYun: $q_{\rm Y01} = \rm{log}_{10}(\frac{\rm{FIR}}{3.75\times 10^{12}\, W \,m^{-2}}) - \rm{log}_{10}(\frac{S_{\rm 1.4\, GHz}}{W\, m^{-2}\, Hz^{-1}})$, where $\rm{FIR}$ is defined as $\text{FIR} \equiv 1.26 \times 10^{-14} (2.58\, S_{60\, \mu m} + S_{100\, \mu m}) \, \text{W m}^{-2}$, and S$_{\rm 1.4\, GHz}$ and S$_{60 ,100 \mu m}$ are the flux densities in units of W \,m$^{-2}$ and Jy respectively. 

We note that these frequencies/wavelengths are frequently studied in the literature in part as free-free contributions typically do not dominate the radio spectrum at 1.4 GHz, and  60-100 $\mu m$ is typically close to the peak of dust emission, where K-corrections may not be as significant (see Figure \ref{fig:sed_demo}). Though, the choice of K-correction may be important for the total FIR-radio correlation, which utilizes the integrated 8-1000$\mu m$ emission. Those quantities may be more sensitive to dust temperature changes that may bias emission significantly at shorter wavelengths and be affected by limited observational constraints at higher $z$ for the `true' dust temperature \citep{liang_dust_2019}, in addition to post-starburst effects \citep{Hayward2014}. We leave the exploration of those effects to future work.

From close inspection of Figure \ref{fig:FRC_all}, a few important results are apparent: 

\textit{1)} The $z \approx$ 0 snapshots, regardless of physics variation, approach the observed values of \qYun, though with larger scatter in the runs including AGN, particularly in \BHkappaconst runs. 

\textit{2)} The FIRE-3 simulations, at all redshifts sampled, regardless of physics variation (\noBHkappaconst, \BHkappaconst, \BHkappavar) roughly lie along a linear FRC at high \Lsixty ($\gtrsim$ 10$^{9.5}$), and shift to a slightly super-linear relation at low \Lsixty ($\lesssim$ 10$^{9.5}$), particularly at higher redshifts ($z \gtrsim$ 1.5). 

\textit{3)} There is little evolution in \qYun with $z$, as galaxies approach the $z=0$ FRC, with the mean value of \qYun evolving by $\sim$1 dex from the $z=4$ value to the \zzero\, value in the \noBHkappaconst, and by $\sim$0.5 dex in the \BHkappaconst and \BHkappavar runs.

The overall scatter in our FRC arises not solely from variation on short ($\lesssim$ 100 Myr) timescale variations in SFH near peak SF for the more massive \texttt{\texttt{m12's}} and \texttt{m13's} \citep{feldmann_colours_2017}, or from late-time bursty star formation/`breathing modes" in the less-massive \texttt{m11's} \citep{muratov_gusty_2015,sparre_starbursts_2017}, but from a combination of such effects and significant galaxy-to-galaxy variation. We confirm this by examining each individual FRC for every simulated galaxy in our mock-observed sample. 

\begin{figure*}\label{fig:q_z_mass}
    \centering
    \includegraphics[width=1.0\textwidth]{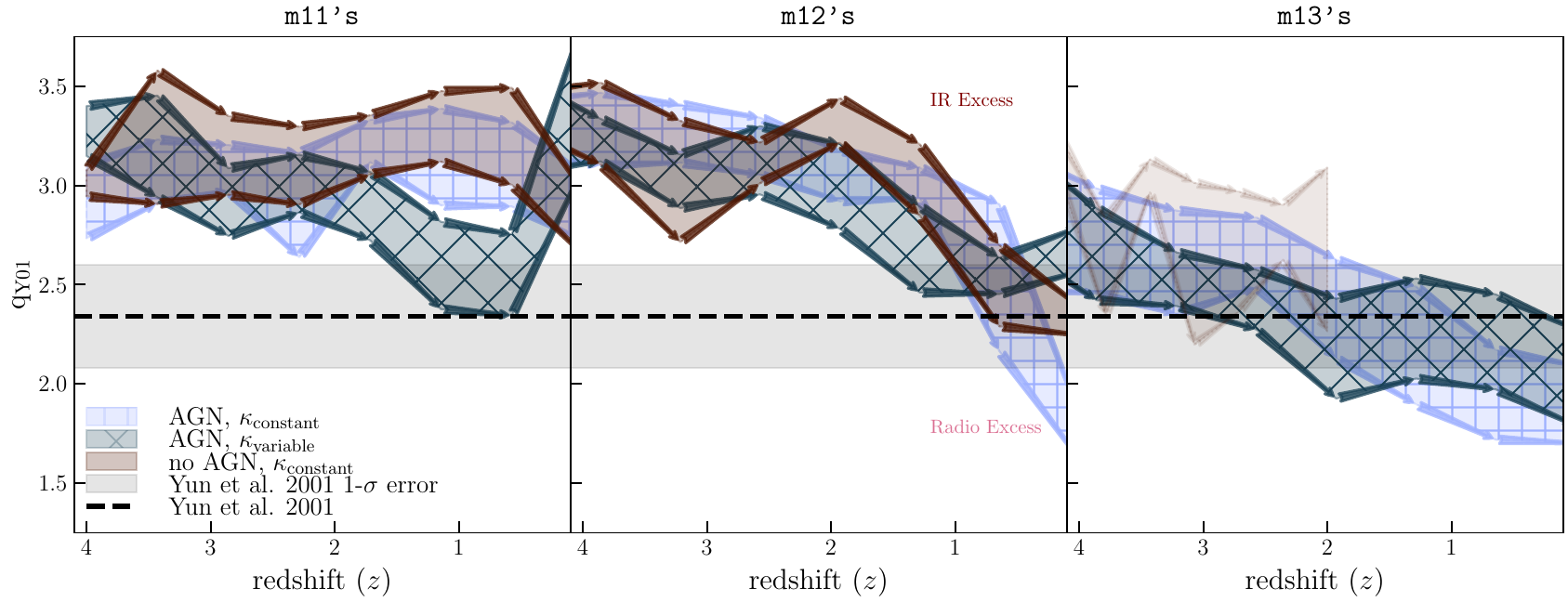}
    \caption{\textit{FRC evolution with redshift by galaxy mass} shown by \qYun vs. $z$ with the same style as Figure \ref{fig:FRC_all}, here grouped by M$_{\rm halo}^{z=0} \sim $ 3 $\times$ 10$^{10} - $ 7 $\times$ 10$^{11}$ M$_{\odot}$ (\texttt{m11's}, \textbf{left}), 7 $\times$ 10$^{11}$ - 1.5 $\times$ 10$^{12}$ M$_{\odot}$ (\texttt{m12's}, \textbf{middle}), and 5 $\times$ 10$^{12}$ - 10$^{13}$ M$_{\odot}$ (\texttt{m13's}, \textbf{right}). A slight systematic difference emerges in \qYun vs. $z$ between the \BHkappaconst and \BHkappavar variants at moderate redshifts, and the \BHkappaconst sample exhibits a larger scatter at $z$=0 with a significant tail of ``radio-excess" objects, particularly at the \texttt{m12} and \texttt{m13} mass scales, owing to longer-lived synchrotron halos arising from Type Ia SNe and BH accretion in quenching galaxies.} 
\end{figure*}

The evolution of the scatter at low-$z$, however, is further modulated by an interplay of AGN and CR feedback, which we further detail in Figure \ref{fig:q_z_mass}, where we break down the total sample in \qYun vs. $z$ space by halo mass groupings and physics variation. Examining Figure \ref{fig:q_z_mass} reveals that even without inserting a `radio-loud' AGN spectrum by-hand, several of our snapshots for the \texttt{m12's} and \texttt{m13's} populate the `radio excess' synthetic-observational parameter space, though not through the mechanism of a `radio-loud' synchrotron jet beamed towards the observer at z $\lesssim$ 1.5. This effect appears to be most pronounced for the \texttt{m12's},  with the systematic difference between the \BHkappaconst and \BHkappavar model variants being most apparent at late times, however it also appears, albeit to a lesser extent, in the evolution of the \texttt{m13's}. The less massive \texttt{m11's}, on the other hand, populate the upper envelopes of the scatter in \qYun vs. $z$ at late times, although examination of individual galaxy FRCs for the more massive \texttt{m11} dwarf galaxies reveals similar behavior to the \texttt{m12's}.

These results reflect the first forward-modeled reproduction of the observed \zzero\, FRC across a wide gamut of galaxy properties using cosmological zoom-in simulations of galaxies with on-the-fly, dynamically evolved CR spectra. Simultaneously, our results illustrate a diversity of paths leading to the \zzero\, FRC across cosmic time owing to CR transport and AGN feedback physics, which we describe in detail below. 

\subsection{Calorimetry \& Conspiracy Continue}\label{sec:CalorimetryandConspiracy}
The quasi-universality of \qYun vs. $z$ indicates a rough balance between the dominant synchrotron intensity-weighted CRe loss rate and UV escape for much of the evolutionary history of our simulated galaxy sample. 
This indicates a simple explanation for the evolution of the FRC from $z=5 \rightarrow 0$: in the standard picture of star-formation-limited CRe and UV calorimetry \citep{voelk_correlation_1989,lacki_physics_2010}, in order to maintain a constant value of \qYun, the galaxies must either be effective CRe and UV-calorimeters (where the synchrotron loss rate is far greater than other relevant loss rates and UV emission is effectively reprocessed) or the synchrotron cooling rate must fall at a similar rate to the UV optical depth (often called `conspiracy' models in the literature). 

We see calorimetry to be generally the case at high \Lsixty $\geq$ 10$^{9.5}$ L$_{\odot}$ (or equivalently at higher galaxy-averaged SFRs, $\Sigma_{\rm gas}$, or $\Sigma_{\rm SFR}$  as often parameterized in one-zone models and other works), irrespective of redshift, with the galaxies with highest \Lsixty corresponding to effective UV and CRe (and CR proton)\footnote{For the simulations in this work, we do not separately evolve secondary-vs-primary electrons in snapshot outputs or include an explicit positron population as in \citep{hopkins_first_2022,ponnada_synchrotron_2024}. Instead secondary CRes (e.g. from pionic decay via CR proton hadronic losses) are directly added to the separately evolved CRe bins in-code. So we cannot directly comment on the secondary contribution, but this channel of CRe production is likely increasingly important at higher $z$ as bremsstrahlung losses become important for CRe \citep{lacki_physics_2010,Werhahn2021}.} calorimeters (also called the `high-$\Sigma_{\rm gas}$ conspiracy' \citealt{lacki_physics_2010}).

However, despite this constancy of \qYun, a large fraction of our synthetic observations populates the `IR-excess' parameter space, particularly for \noBHkappaconst runs at higher redshifts ($z \gtrsim 1.5$). This does not indicate excessive FIR emission, but rather a greater super-linear deviation from radio continuum-SFR calorimetry \citep{condon_radio_1992} than from FIR-SFR calorimetry \citep{kennicutt_star_1998} for our sample between $z\sim5\rightarrow1.5$, and so are better understood as `radio-dim' objects at higher redshifts. 

\begin{figure}
    \centering
    \includegraphics[width=0.5\textwidth]{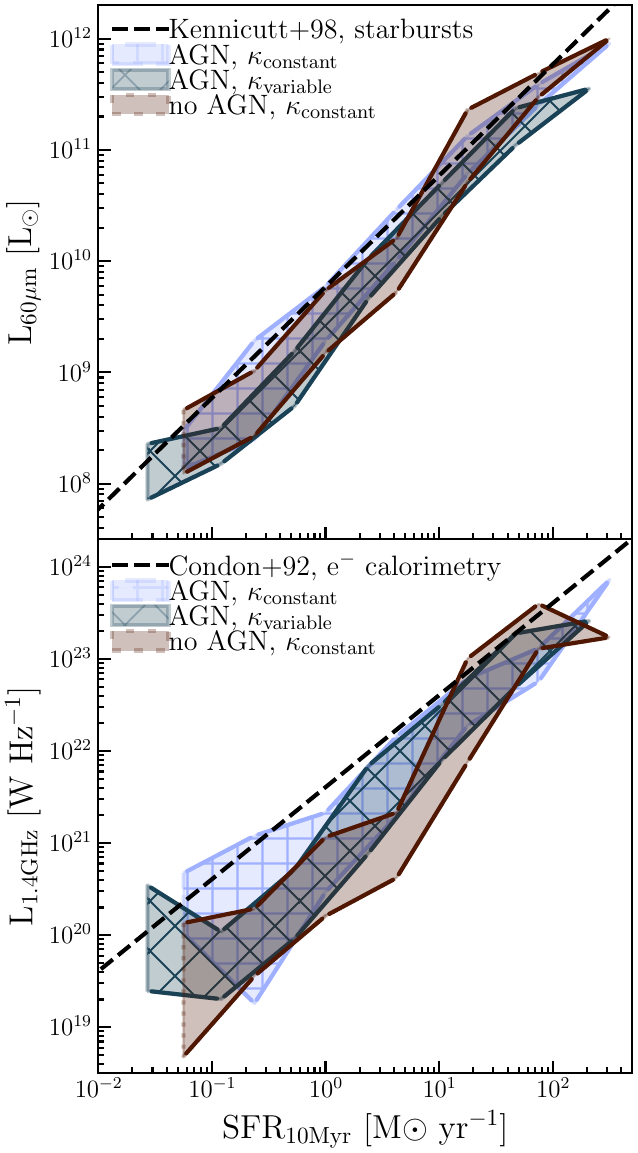}
    \caption{\textit{Calorimetry \& Conspiracy:} \Lsixty \,(\textbf{top}) and \Lradio\,(\textbf{bottom}) vs. 10-Myr averaged SFRs for snapshots from with \noBHkappaconst, \BHkappaconst, and \BHkappavar\, in brown with no hatching, light blue with square hatching, and navy with diamond hatching respectively. Shaded regions show the 32-68 percentile (approximate $\sim 1\sigma$) confidence intervals and line segments demarcate bins. Expected calorimetric relations for the IR with SFR from starbursts \citep{kennicutt_star_1998} and CRe calorimetry at 1.4 GHz \citep{condon_radio_1992} are shown in black dashed lines. Across all physics variations, our simulations generally approach calorimetric expectations at high \Lsixty (\Lsixty $\geq$ 10$^{9.5}$ L$_{\odot}$), with super-linear relations at low \Lsixty (\Lsixty $\leq$ 10$^{9.5}$ L$_{\odot}$) (or low $\Sigma_{\rm gas}$) owing to lower UV optical depths and high $\tau^{-1}_{\rm diff}/\tau^{-1}_{\rm synch}$ in weaker on average \textbf{B$_{\rm \bot}$}. At low SFRs ($\leq$ 0.5 $\rm{M_{\rm \odot}} yr^{\rm -1}$), our runs with AGN more often exhibit `super-calorimetric' behavior in \Lradio\, with scatter extending beyond the `calorimetric' relation, particularly at SFRs $\lesssim$ 0.1 $\rm{M_{\rm \odot}} yr^{\rm -1}$ owing to late-time contributions of CRs from Type Ia SNe and BH accretion in quenching galaxies.} 
    \label{fig:calorimetry}
\end{figure}

We confirm this by examining the 10 Myr-averaged SFRs at each snapshot in comparison to the literature calorimetric relations of \citet{condon_radio_1992,kennicutt_star_1998} for all of our synthetic observations in Figure \ref{fig:calorimetry}. The larger relative loss of radio-emitting CRe compared to UV-emitting photons arises from a relatively ubiquitous prediction for our low-\Lsixty\, simulated galaxies across our physics model variations, which comprise a larger fraction of our sample with increasing redshift (as \texttt{m11's} and \texttt{m12's} have rising star-formation histories (SFHs) with time),  wherein the dominant loss rate of synchrotron-emitting CRe increases faster than the dust opacity to UV radiation decreases.  

This is not wholly unexpected from observations, as recent studies exploring the L$_{\rm IR}$-SFR \citep{bonato_correlations_2024} have found a transition to increasingly super-linear L$_{\rm IR}$ $\propto$  SFR$^{\alpha}$ at low total IR 8-1000$\mu$m luminosities (L$_{\rm TIR} \lesssim 10^{10} \rm{L}_{\rm \odot}$ or SFRs $\lesssim 1\, \rm{M}_{\rm \odot}\, \rm{yr}^{-1}$, with $\alpha \gtrsim$ 1.25 and increasing with lower L$_{\rm TIR}$, as the obscured SFR fraction decreases at lower galaxy masses \citep{Hayward2014,whitaker_constant_2017}. Similarly, stacking analyses of \, spiral galaxies have found even steeper best-fit 1.4 GHz radio-UV+TIR-estimated SFRs for galaxies with M$_{\ast} \leq$ 10$^{\rm 10.5}$ M$_{\rm \odot}$, where L$_{\rm 1.4\, GHz}$ $\propto$  SFR$^{\beta}$, where $\beta \sim$ 1.5. Together, these trends naturally explain the modest hints of non-linearity in the FRC at low luminosities, with \qYun increasing with decreasing \Lsixty \citep{yun_radio_2001,bell_estimating_2003,matthews_cosmic_2021}, and are in agreement with predictions from previous non-calorimetric one-zone steady-state models \citep{lisenfeld_quantitative_1996,lacki_physics_2010}.

These trends are consistent with the behavior predicted by our simulations, and explained by examination of the relevant CRe cooling rates. The IC and synchrotron cooling rates are given by the following:

\[
\tau^{-1}_{\rm IC,\, synch} = (4/3)\,\sigma_{\rm T}\,\gamma_{\rm CRe}^{2}\, c\,(u_{\rm rad} , u_{\rm B})/E_{\rm CRe}
\]

\citep{Rybicki}, where $\sigma_{\rm T}$ is the Thomson cross-section, $\gamma_{\rm CRe}$ is the CRe Lorentz factor, E$_{\rm CRe}$ is the characteristic energy of CRes emitting at the observing frequency\footnote{Though note, in detail the emission at a given frequency arises from a \textit{distribution} of energies at a given \textbf{B$_{\bot}$}. However, this approximation is satisfactory for the order-of-magnitude comparisons here, despite the detailed forward-modeled emission utilizing multi-bin spectra.} for a given gas cell's \textbf{B$_{\rm \bot}$} (typically $\sim$ 0.5-30 GeV at 1.4 GHz), $u_{\rm rad}$ is the radiation energy density given self-consistently from summing all bands followed in our in-code radiation-hydrodynamics approximation in addition to the un-attenuated CMB, and $u_{\rm B}$ is the magnetic field energy density from our explicitly-evolved \textbf{B}. The effective diffusive escape time can be computed as follows:

\[
\tau^{-1}_{\rm diff} \sim v_{\rm stream}^{\rm iso}/\ell_{\rm CRe}
\]

where $v^{\rm iso}_{\rm stream} \sim \kappa^{\rm iso}_{\rm eff}/\ell_{\rm CRe}$ is the isotropically-averaged effective CRe streaming speed (determined by the CR fluxes in-code, $v_{\rm stream} \sim \langle {\bf F}_{\rm CRe}\cdot \hat{g} / e_{\rm CRe} \rangle $ where $\hat{g}$ is the direction of $\nabla e_{\rm CRe}$), around the CRe energies which matter for the synchrotron losses of interest and $\ell_{\rm CRe} \equiv e_{\text{CRe}}/\nabla_{\parallel} e_{\text{CRe}}$ is the CRe pressure scale length. For simplicity, since this is order-of-magnitude anyways, we adopt a median $\ell_{\rm CRe} \sim {\rm kpc}$ for our analysis, but using a variable number like the local disk scale length or galaxy size or median of different cell gradients in the galaxy gives similar results.

Comparing the synchrotron-intensity-weighted averages of the IC and diffusive loss rates for each of the synthetically-observed snapshots shows a general trend of IC loss rates being roughly similar to the synchrotron loss rates at all $z \lesssim 2$, with $\langle\tau^{-1}_{\rm IC}\rangle_{I_{\rm \nu}}$ $\sim$ 0.1-3 $\langle\tau^{-1}_{\rm synch}\rangle_{I_{\rm \nu}}$, rarely exceeding $\sim$ 10, broadly indicating equipartition between u$_{\rm rad}$ and u$_{\rm B}$ in synchrotron-emitting gas at lower redshifts, but $\langle\tau^{-1}_{\rm IC}\rangle_{I_{\rm \nu}}$ $\sim$ 10$^{-0.5}$ - 10$^{2}$ $\langle\tau^{-1}_{\rm synch}\rangle_{I_{\rm \nu}}$ at $z \gtrsim 2$. Surprisingly, even at higher redshifts where u$_{\rm rad, CMB} = 0.26\, (1+z)^{4}\, \rm{eV\, cm^{-3}}$ becomes large, the average IC loss rate can occasionally be comparable to that of synchrotron losses in dense, strongly magnetized, high synchrotron-emissivity gas! Meanwhile, $\langle\tau^{-1}_{\rm diff}\rangle_{I_{\rm \nu}} \gtrsim$ 10-10$^{2}$ $\langle\tau^{-1}_{\rm synch}\rangle_{I_{\rm \nu}}$ for low \Lsixty\, snapshots further off the FRC at all $z$, but with $\langle\tau^{-1}_{\rm diff}\rangle_{I_{\rm \nu}}$ $\sim$ 0.1-10 $\langle\tau^{-1}_{\rm synch}\rangle_{I_{\rm \nu}}$ at high \Lsixty, maintaining standard arguments of calorimetry and conspiracy at high and low $\Sigma_{\rm gas}$ respectively. In short, both IC losses and diffusive CRe escape play important roles at $z \gtrsim 2$ in shaping the FRC at the stellar masses/gas surface-densities sampled here, resulting in `radio-dim' snapshots, whereas at $z \lesssim 2$, diffusive escape becomes the primary competing loss term to synchrotron losses.

\subsection{Beyond Calorimetry \& Conspiracy: `Hooks', `Lines', and `Sinkers'}
\begin{figure*}
    \centering
    \includegraphics[width=1.0\textwidth]{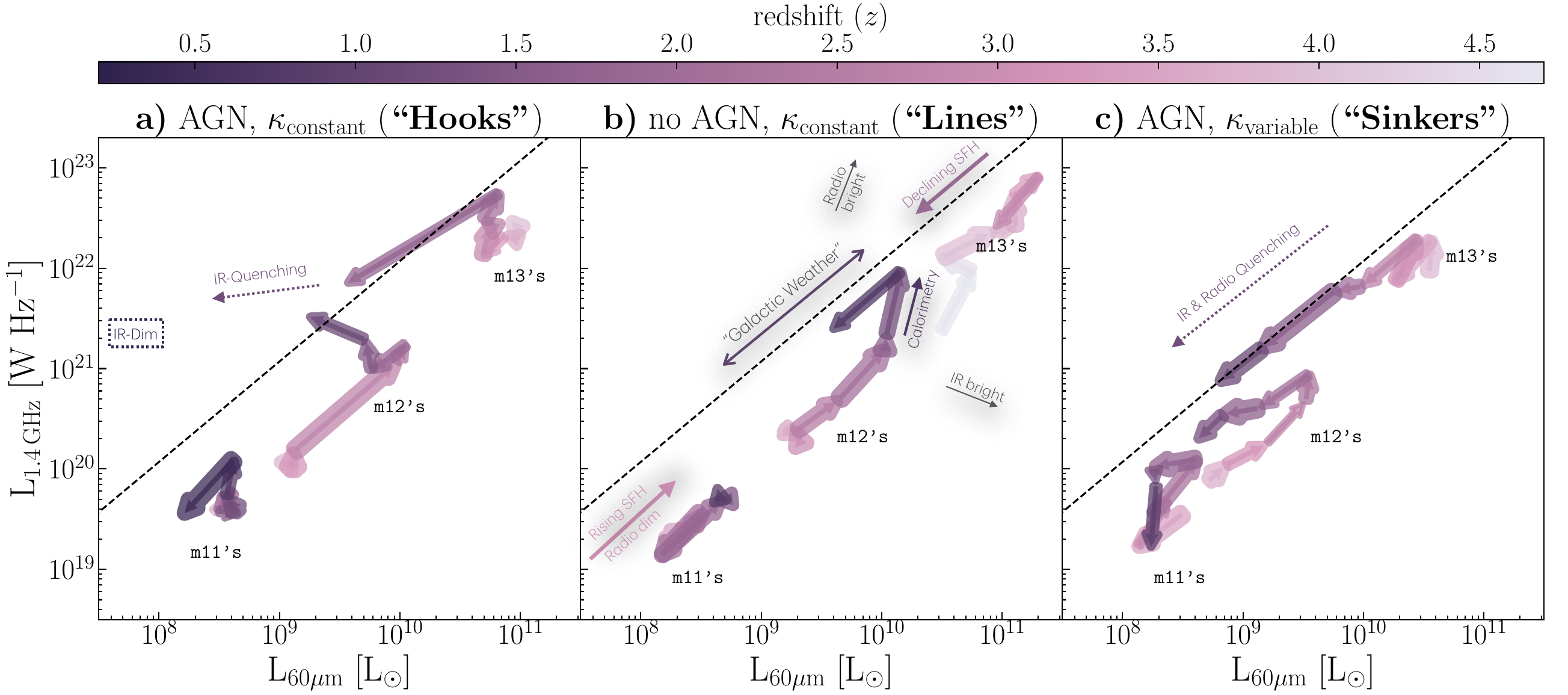}
    \caption{\textit{FRC ``tracks'' on \Lradio\, vs. \Lsixty\, color coded by redshift} for \textbf{a)} \BHkappaconst (\textbf{left}), \textbf{b)} \noBHkappaconst (\textbf{center}), and \textbf{c)} \BHkappavar physics variations (\textbf{right}), with the \citet{yun_radio_2001} \zzero\, relation (black dashed line). The thickness of each arrow, colored by median redshift, qualitatively shows the relative 1$\sigma$ scatter in a given redshift bin for the galaxies, which are mean averaged together by log$_{10}$(\Lsixty) and log$_{10}$(\Lradio) by mass groupings M$_{\rm halo}^{z=0} \sim $ 3 $\times$ 10$^{10} - $ 7 $\times$ 10$^{11}$, 7 $\times$ 10$^{11}$ - 1.5 $\times$ 10$^{12}$, and 5 $\times$ 10$^{12}$ - 10$^{13}$ M$_{\odot}$ (\texttt{m11's}, \texttt{m12's}, \texttt{m13's}). The different physics variations show different evolutionary paths with respect to $z$ -- \textbf{solid annotations} with shadowed tint in the middle panel highlight physics operating in all models while \textbf{dotted annotations} highlight the characteristic differences from the \noBHkappaconst scenario, driven by AGN feedback + CR transport physics. In \BHkappaconst models, \texttt{m12's} and \texttt{m13's} typically evolve parallel to the observed \zzero\, FRC, tracking smoother, rising/constant SFHs before quenching at late times owing to AGN feedback, subsequently moving perpendicular to and `hooking' the FRC, becoming `IR-dim' objects with long-lived synchrotron contributions from Type Ia SNe and BH accretion. In \noBHkappaconst models, \texttt{m12's} and \texttt{m13's} typically rise up to the FRC `line' tracking their SFHs, before exhibiting scatter slightly \textit{along} the FRC owing to galaxy-to-galaxy scatter and late-time `galactic weather' effects. \BHkappavar models show a different here evolutionary track along the FRC, with \texttt{m12's} evolving up to and along the \zzero\, FRC prior to `sinking' $\sim$1 dex in \Lsixty, and \texttt{m13's} steadily `sinking' by $\sim$2 dex in \Lsixty from $z=5$ to $z=0$. For all three physics models, the averaged m11 tracks exhibit noisier variations due to burstier SFHs and larger galaxy-to-galaxy variations at this mass scale, though the more massive \texttt{m11's} exhibit qualitatively similar behavior to the \texttt{m12's} and \texttt{m13's}.} 
    \label{fig:tracks}
\end{figure*}

In Figure \ref{fig:tracks}, we show the differing characteristic evolutionary paths of galaxies of the \noBHkappaconst, \BHkappaconst, and \BHkappavar physics variations, with annotations indicating general behaviors characterizing all models shown in the \noBHkappaconst case (\textit{middle panel, solid annotations}). Here, we average groups of galaxies by $z=0$ halo mass corresponding to \texttt{m11's}, \texttt{m12's}, and \texttt{m13's} binning the synthetic observations meeting our selection criteria by redshift. Thus, the tracks detail temporal as well as galaxy-galaxy scatter. 

Upon examining each individual FRC track and comparing to average trends shown here, we are able to characterize some general behaviors of each physics model. While there are some outlier galaxies within each physics variation of our sample, the general tracks of `hooking' across the FRC to become relatively `radio-excess' objects at late times in the \BHkappaconst case, gradually moving up to the observed FRC with late-time scatter primarily along `lines' in the \noBHkappaconst case, and following similar early-time behavior before `sinking' further down the FRC line in the \BHkappavar case appear to be robust trends, and qualitatively explain the differences in the normalization and scatter of \qYun\, presented in Figure \ref{fig:q_z_mass} as a function of $z$. We expand upon and discuss the physical mechanisms driving these ``tracks'' and the late-time ($z \lesssim 1.5$) below.

\subsubsection{`Hooks': The FRC of \BHkappaconst}\label{sec:hooks}
From Figures \ref{fig:FRC_all}, \ref{fig:q_z_mass}, and \ref{fig:tracks}, we see that the general behavior of the \BHkappaconst model variation is to maintain a roughly constant \qYun with redshift but with increasing scatter at late times ($z\lesssim1.5$), with the largest $\sim$ 1$\sigma$ confidence interval, indicating a population of `radio-excess' objects in our sample. Within each grouping of galaxies by halo mass, the average trend is to `climb' up towards the FRC from early to late times as galaxies build up stellar mass with rising SFHs, often before `hooking' across the observed \zzero\, FRC with nearly perpendicular tracks, in many cases becoming `radio-excess' objects at $z\sim 0$, ubiquitously for the \texttt{m12's} and \texttt{m13's}. While not as evident in Figures \ref{fig:q_z_mass} and \ref{fig:tracks}, we see similar behavior for the more massive \texttt{m11's} at the $\sim$ 3-5 $\times 10^{\rm 11}\, \rm{M}_{\odot}$ mass scale, which we show an example of in Figure \ref{fig:m11f}.

This `hook' pattern posits a somewhat obvious a posteriori, but previously un-explored progenitor scenario of `radio-excess' objects across a broad range of \Lsixty\, (from 10$^{7.5}$ - 10$^{11.5}\, \rm{L_{\odot}}$): \textbf{`radio-excess' objects can in fact simply be `IR-dim,' rather than intrinsically radio-bright}. Recall that in our synthetic observations here, we do not insert by hand any hard radio spectrum or template AGN spectrum, and remove from our sample snapshots with accretion rates which would noticeably pollute the \Lsixty-\Lhund\, luminosities. Thus, these are solely the \textit{indirect} effects of AGN on the observables --  meaning the evolution of \qYun seen here is the end product of generic galaxy formation physics related to cosmological galaxy growth and quenching due to AGN+CR feedback, of which the exact sub-grid accretion disk physics \citep{hopkins_2024_forged_in_fire_1} and jet-launching criteria \citep{su_unraveling_2023} remain uncertain, and the emergent grid-scale transport rates of CRs from micro-physical, unresolved phenomena, of course, remains orders-of-magnitude uncertain in various phases of the ISM and CGM \citep{hopkins_testing_2021,hopkins_standard_2022,hopkins_standard_2022,kempski_reconciling_2022,butsky_constraining_2023,thomas_cosmic-ray-driven_2023,butsky_galactic_2024}.  

Rather than tracing sparse star formation in any meaningful sense, \Lradio\, in these quenching objects at late times traces CR contributions from Type Ia SNe, which dominate at sSFRs $\lesssim$ 10$^{-11}$, as well as injection from episodic BH accretion at late times. This is a strictly `non-calorimetric' effect with regards to star-formation, which is pronounced in \qYun owing to the cessation of young star formation, which \Lsixty-\Lhund\, is heavily sensitive to.

\subsubsection{`Lines': The FRC of \noBHkappaconst}\label{sec:lines}
The behavior of the \noBHkappaconst physics variation in Figure \ref{fig:tracks} is largely summarized in Section \ref{sec:CalorimetryandConspiracy}. Our simulated galaxies run with this physics model, are most similar in approach to other galaxy simulations including CR-MHD in the literature \citep{Werhahn2021,werhahn_cosmic_2021,werhahn_cosmic_2021b,pfrommer_simulating_2022,farcy_radiation-magnetohydrodynamics_2022,thomas_cosmic-ray-driven_2023,rodriguez_montero_impact_2024, martin-alvarez2024}, up to the difference of explicit, dynamic, on-the-fly evolution of CR(e) spectra here, and offer a useful point of comparison. 

Across the mass range sampled here, our \noBHkappaconst galaxies largely rise up to the \zzero\, FRC tracing their rising/constant SFHs with time, with lower-mass, low \Lsixty\, ($\leq$ 10$^{9.5}$ L$_{\odot}$) galaxies having larger values of \qYun associated with super-linear deviations from L$_{\rm IR}$-SFR and \Lradio-SFR correlations, and more-massive, higher \Lsixty\, snapshots approaching calorimetry and subsequently the FRC. 

There is significant snapshot-snapshot scatter in our \texttt{m11's} and \texttt{m12's} at late times for a given galaxy, and considerable galaxy-to-galaxy variation contributing to the scatter, but it largely manifests \textit{along} the relation, with `galactic weather' of short-lived starbursts, outflow events, and late-time mergers tending to move galaxies parallel to the FRC rather than `off' the relation. This is the primary cause of the `downward' arrow pointing for the \texttt{m12's} ``track" in Figure \ref{fig:tracks}, which is of a qualitatively different nature than the `sinking' described in the next subsection (c.f. Figure \ref{fig:cartoon}).

We note that the \texttt{m13's} included in this model variant sample are not run to low-$z$ (typically halted at around $z\sim2$, as indicated in Figure \ref{fig:q_z_mass}) as without AGN feedback, galaxies at this mass-scale fail to quench star formation and consequently over-/under-shoot  low-$z$ constraints by upwards of an order-of-magnitude, often becoming too dense in their centers to continue running the simulation forward in time \citep{byrne_formation_2023}. However, the high-$z$ \texttt{m13} snapshots represent an interesting sample of brighter, denser, and actively starbursting galaxies near cosmic noon, and so we include them here. Even in this more extremal end of $\Sigma_{gas}$, we find that the FRC holds, indicating that despite relatively higher bremsstrahlung, Coulomb and IC losses in this regime, secondary contributions from CR proton spallation `save' the FRC at high-$\Sigma_{gas}$.

\subsubsection{`Sinkers': The FRC of \BHkappavar}\label{sec:sinkers}

The third pathological behavior we observe in our physics variation of \BHkappavar is similar in many aspects to the \BHkappaconst evolutionary path described in \ref{sec:hooks}, though with important qualitative differences. After entering our mock-observable parameter space from $z > 5$, the \texttt{m13's} appear to `sink' down along the \zzero\, FRC by $\gtrsim 2$ dex in \Lsixty\, and \Lradio, with low-$z$ snapshots contributing to the sample at low-\Lsixty, filling in the upper envelope of the scatter in Figure \ref{fig:FRC_all}. From examining the individual `tracks' of each galaxy, we note that while there are some low SFR snapshots excursing into the `radio-excess' parameter space, these occur less often than in the \BHkappaconst case, and often stabilize at later times back onto the \zzero\, FRC as evinced by the pointings of the averaged arrows in Figure \ref{fig:tracks}. 

The \texttt{m12's} in this model variant exhibit similar behavior to the \BHkappaconst runs as well, but also `sink' at late times owing to quenching, much alike the \texttt{m13's}, populating the low-\Lsixty\, parameter space at z $\sim$ 0. Although the \texttt{m11's} exhibit qualitatively more scatter, it is primarily coherent along the FRC, indicating similar regulation of \Lsixty\, and \Lradio\, during sporadic burst-quench cycles via effective loss of dust opacity in addition to relatively faster diffusive escape of CRs, especially for the more massive \texttt{m11's} (see Figure \ref{fig:m11f}). The averaged \texttt{m11's} track, however, becomes more `radio-dim' at the latest times, indicating the balance of CRe diffusive loss vs. UV opacity is skewed at this mass scale, though this may be partly due to our use of a constant D/Z value of 0.4. For lower D/Z at the lowest masses, the `sinking' behavior would then appear more similar to the \texttt{m12's} and \texttt{m13's}. 

This explains the relatively smaller scatter of the \BHkappavar model variation at late times in comparison to the \BHkappaconst runs; as the parallel diffusivity is allowed to vary with local plasma properties in the $\kappa_{\rm var}$ model, this allows for larger effective volume-averaged transport speeds, limiting the presence of longer-lived synchrotron emission in `IR-dim' galaxies.

Recall that if we assume the dynamical equations for the CR flux and scattering rates have all reached steady-state in the diffusive limit, the amplitude of gyro-resonant scattering modes $\delta B^{2}(\lambda \sim r_{g}) \sim u_{\delta B}^{\rm gyro}$ is set by balance between some source terms $S_{\pm}$ and damping terms $\Gamma_{\pm}$, so $\dot{u}_{\delta B}^{\rm gyro} \sim +S_{\pm} - \Gamma_{\pm} u_{\delta B}^{\rm gyro} \rightarrow 0$, i.e.\ $\delta B^{2}/B^{2} \rightarrow S_{\pm}/\Gamma_{\pm} u_{\rm B}$, with the emergent parallel diffusion coefficient being $\kappa_{\|} \sim v_{\rm cr}^{2}/3\bar{\nu}_{\rm s} \sim (v_{\rm cr}^{2} / \Omega_{\rm cr})\,(B^{2}/\delta B^{2}) \propto S_{\pm}^{-1}$ (with $\Omega_{\rm cr}$ the gyrofrequency; \citealt{zweibel_microphysics_2013}). 

Taking this intuition and the source term used in the $\kappa_{\rm var}$ models (see Section \ref{sec:sample}), in conjunction with standard scalings for ``turbulent," Linear Landau, or collisionless damping for $\Gamma_{\pm}$ leads to an approximate scaling of $\kappa_{\|} \propto \textbf{B}^{-\xi}$, where $\xi$ is some weak power-law index ($\xi \sim$ 0.2-0.5), of particular interest here for synchrotron calculations, notwithstanding weak dependencies on other plasma properties marginalized over. Essentially, denser, more strongly-magnetized regions of the galaxy tend to have higher turbulent dissipation rates, which drives stronger CR scattering (more CR confinement).

Despite this inverse scaling of $\kappa_{\|} \propto \textbf{B}^{-\xi}$, these simulations tend to `sink' down the FRC rather than becoming `radio-excess' objects as they quench. This is somewhat non-intuitive --  if CRes effectively spend more time in strongly magnetized gas, one might expect that these simulations would be more `radio-excess' on average. 

However, the scattering properties of CRs are better understood in the full context of the multi-phase ISM/CGM. If CRes primarily emit in dense, magnetized, neutral gas \citep{ponnada_synchrotron_2024,martin-alvarez2024} as expected from the power-law relationship of $\textbf{B}-\rho^{0.4-0.6}$ \citep{Tritsis2015,Ponnada2022}, and this gas is enveloped in a uniform bath of more diffuse phases where $\kappa_{\rm eff}$ is higher on average, the \textit{global} synchrotron properties will be set simply by the `boundary' condition of entering regions of approximate or total synchrotron calorimetry. So even if $\kappa_{\|} \rightarrow 0$ in strongly magnetized gas in the most extreme example, the rate limiting step for producing synchrotron emission is the scattering of CRes into such high-\textbf{B} regions, which on average is \textit{lower} in this type of model, wherein $\kappa_{\rm eff}$ is higher outside these regions and CRs may more quickly escape the scattering halo of characteristic size $\sim$ 10 kpc in which they have an order unity probability of re-entering the relatively synchrotron-bright gas disk/bulge \citep{hopkins_testing_2021}.

\subsubsection{Morphological Variation Across `Tracks'}

\begin{figure*}
    \centering
    \includegraphics[width=1.0\textwidth,height=1.0\textheight,keepaspectratio]{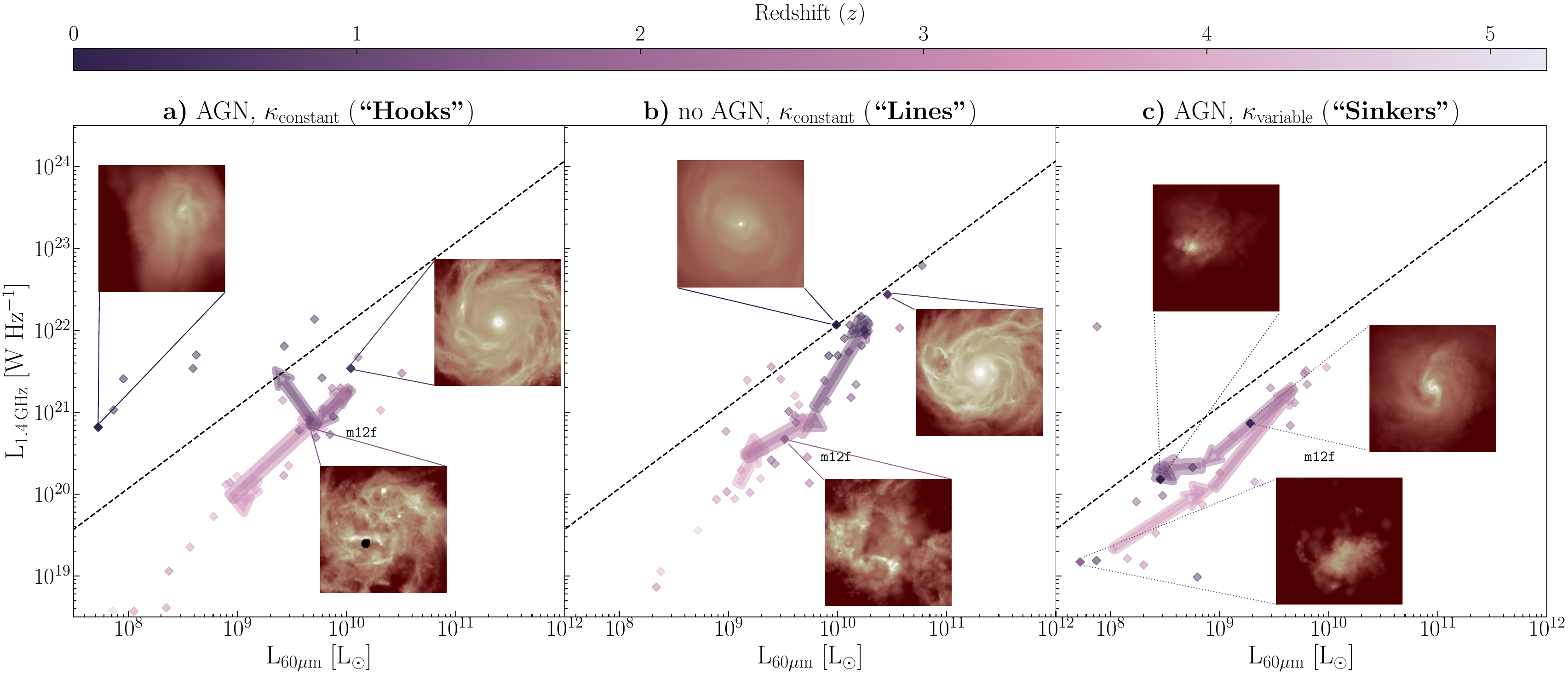}
    \caption{\textit{Morphological evolution along FRC ``tracks''}, here shown for an individual halo, \texttt{m12f}, as a representative case study for \textbf{a)} \BHkappaconst (\textbf{left}), \textbf{b)} \noBHkappaconst (\textbf{center}), and \textbf{c)} \BHkappavar physics variations (\textbf{right}), with individual snapshots shown as points (pluses, circles, diamonds), along with the \citet{yun_radio_2001} \zzero\, relation (black dashed line).  Properties for the arrows are the same as in Figure \ref{fig:tracks}. Insets show 1.4 GHz images in the central 25 kpc for three snapshots sampling early to late-time evolution in each case, highlighting characteristic underlying behavior for each model class' ``track" described in Sections \ref{sec:hooks} - \ref{sec:sinkers}.} 
    \label{fig:cartoon}
\end{figure*}

The different `tracks' outlined above trace different stages of galaxy evolution, and thus different gas morphological properties. In particular, `IR-dim' objects often display morphological disturbance in gas phase, with more lenticular-like synchrotron morphology and extended diffuse emission. We illustrate such characteristic morphological evolution in each track for one of our $\sim L^{\ast}$ halos, \texttt{m12f}, as a representative example in Figure \ref{fig:cartoon}.

 The progression of simulations along each track becomes apparent from visual examination of Figure \ref{fig:cartoon}: in each of our physics models, galaxies largely `climb' up the FRC from irregular gas morphologies where they are `radio-dim' on average due to stronger losses of CRes. The late-time morphological evolution, however, differs between the no AGN and AGN runs, with \noBHkappaconst runs at this mass scale maintaining star-formation and extended, structured gas distributions. The scatter subsequently manifests along the relation from late-time mergers and feedback episodes reflected by the more modest changes in gas morphology.

\BHkappaconst and \BHkappavar runs contrast this with irregular, gas-deficient morphologies after quenching star formation, with $\kappa_{\rm const}$ runs more often becoming `IR-dim' at late times by maintaining more extended, bright synchrotron emission from CRe produced by intermittent BH accretion and Type Ia SNe, while $\kappa_{\rm var}$ runs `sink' down the correlation on average owing to faster CR diffusion.

\section{Discussion and Conclusions}\label{sec:discussion}

The globally-averaged properties of our synthetically observed sample here generally verify the predictions of non-calorimetric one-zone models \citep{lisenfeld_quantitative_1996,lacki_physics_2010} while evolving the details of stellar feedback, thermochemistry and ISM phase structure, magnetic fields, CR transport and its back-reaction on gas. This is  all done in a cosmological context from relatively well-understood physics undergirding stellar evolution and gas cooling/heating with varied assumptions regarding far-less understood CR transport, here treated using a simple, constant in space and time model for the scattering rate $\nu_{\rm CR}$ as well as a model in which $\nu_{\rm CR}$ varies by orders of magnitude according to local plasma properties. This verification is particularly true of our \noBHkappaconst sample, which are most comparable to the aforementioned works when considering galaxy-averaged quantities in the context of the global FRC as we do here. 

This simulation and synthetic observational effort illuminates physics driving the FRC in detail, which require this approach --  the analysis of our \noBHkappaconst sample highlights how galaxies evolve with redshift to the \zzero\, FRC and how relatively simple assumptions of constant power-law scalings for the $\kappa_{\|}$ motivated by empirical Milky-Way LISM constraints can reproduce the properties of the FRC reasonably well across a diversity of galaxy properties. Moreover, the detailed forward modeling of the emergent observables allows us to forego any implicit assumptions regarding dust optical depth to UV photons, which by construction constrains the predicted FIR luminosities, or resort to re-normalization schemes of spectra solved via steady-state assumptions in conjunction with a diffusion-loss treatment for CR transport, as has been invoked in other simulation works \citep{Werhahn2021,pfrommer_simulating_2022}. These assumptions which may significantly influence the predicted radio continuum synchrotron emission, particularly in galaxies with gas-dense central bulges or generally high gas surface densities \citep[see][for a detailed discussion]{ponnada_synchrotron_2024}.

Beyond calorimetry and conspiracy, our \BHkappaconst and \BHkappavar sample show how the FRC's scatter is shaped by galaxy evolution processes, in particular the role of quenching and the indirect effects of AGN on observables via feedback. Our finding that `radio-excess' objects can instead simply be `IR-dim' objects poses consequences for the `small' scatter often quoted with regards to the FRC. If these objects are simply excluded from observational samples looking to characterize the FRC as `radio-excess,' even if they have low-level star formation, the resulting scatter will by construction be smaller due to sampling bias.

In addition to this, studies aiming to infer AGN jet power from `radio-excess' objects -- especially if these are not high-accretion rate objects (as our selection criteria aim to exclude), the subsequent radio luminosity may have little to do with the intrinsic jet power of the AGN, and can even be dominated by Ia contributions in many cases! Indeed, lacking a better alternative observational metric, state-of-the-art observational studies utilize the excess from the FRC in order to constrain `jet' power and its associated correlations with host galaxy properties \citep{jin_host_2024}. Curiously, taking these nominal constraints as informative regarding the `jet' often leads to finding that the radio-AGN host galaxy quenching ages ($\gtrsim$ 3 Gyr) estimated from stellar mass assembly histories are often poorly correlated with even the oldest jet `ages' from dynamical or spectral age estimations ($\sim$ 1 Gyr; \citealt{Turner2018}). This serves as an additional point of support for these lower-luminosity `radio-excess' objects more physically being `IR-dim,' bearing similarity to some of our \BHkappaconst \texttt{m13's} at late times. This motivates further study of these types of simulations capable of evolving CR dynamics from AGN and stellar sources to aid in disambiguating the nature of observed `radio-excesses' from the FRC in radio AGN studies.

Comparing the \BHkappaconst and \BHkappavar physics variations, it is quite remarkable that the global properties of the FRC can be robustly reproduced despite $\kappa_{\rm eff}$ varying by \textit{orders of magnitude} in the $\kappa_{\rm var}$ model. While there are notable differences discussed prior regarding their scatter and evolution along diverging tracks, it appears that the global FRC may not precisely constrain the transport properties of CRs, at least for the empirically-motivated models explored here. 

However, the differences in gas morphology and subsequent global FRC `tracks' at late times hints at the possibility to constrain CR transport using the spatially-resolved FRC \citep{murphy_initial_2006} as observational works have begun to attempt \citep{2019A&A...622A...8H}. Between our \BHkappaconst and \BHkappavar models, even if bulk FRC properties are similar on aggregate, the exact distribution of synchrotron emission with respect to FIR emission will differ. We caution, however, that this may not immediately make clear how the diffusion properties in different gas phases translate to the radio properties, as we highlight in Section \ref{sec:sinkers}, due to the non-linear nature of CR transport \citep[see also][for further caveats of physically-motivated CR scattering models]{Ponnada2024b}. 

As CRs from BHs in our simulation are injected at large radii compared to the scales of the BH accretion disk, they are more analogous to stellar CRs in these models as their injection and diffusion properties are set by the ISM and CGM. This indicates that BH contributions may potentially explain some of the radio-excesses observed in the spatially resolved FRC at low FIR luminosities \citep{murphy_initial_2006}.

\section{Summary and Future Work}
In this paper, we have presented end-to-end forward modeled synthetic observations of FIR and radio continuum emission from cosmological zoom-in simulations with ``live" CR proton and electron spectra. We forward model the FRC for galaxies from the $M_{\rm halo}^{z=0} \sim$3$\times 10^{10}$ - 1$\times 10^{13}$ $M_{\rm \odot}$ from $z = 5$ to $z = 0$ with three different physical models: one without AGN feedback and a constant-in-space and time treatment for the CR scattering rate, and two with multi-channel AGN feedback and two different models for the CR scattering rate.

Our results can be summarized as follows:
\begin{itemize}
    \item All physics variations explored herein generally reproduce the \zzero\, FRC, but with larger scatter in runs with AGN. Our forward-modeled FRC is linear at high \Lsixty\, $\gtrsim$ 10$^{9.5}$ L$_{\odot}$, but exhibits super-linearity at low \Lsixty\, $\lesssim$ $10^{9.5}$ L$_{\odot}$. 
    \item This FRC is maintained by standard arguments of `calorimetry' and 'conspiracy' at high and low $\Sigma_{\rm gas}$ respectively, with diffusive escape being the primary competing loss process to synchrotron-emitting gas at $z \lesssim 1.5 $ and both IC and diffusive losses being important at higher $z$.
    \item How the late-time scatter of the FRC is shaped is sensitive to the $indirect$ effects of AGN feedback as well as CR transport on the emergent observables. These effects create characteristic evolutionary ``tracks'' with galaxies evolving up to the \zzero\, FRC before exhibiting differing behavior due to the physics variations, which aid in understanding the nature of `outliers' to the FRC.
    \item Namely, galaxies run with our \noBHkappaconst model typically evolve up to and along `lines' parallel to the \zzero\, FRC at late times, with scatter owing to "galactic weather" effects.
    \item This differs from our \BHkappaconst model, more massive dwarf galaxies as well as $\sim L^{\ast}$ and massive elliptical galaxies evolve quasi-perpendicularly to the FRC as galaxies begin to quench, creating `hook' patterns and becoming `IR-dim' objects at late times, filling in the `radio-excess' space with late-time contributions from BH accretion and Type Ia SNe. In our \BHkappavar models, however, these same galaxies steadily `sink' down along the FRC, exhibiting qualitatively less late-time scatter as a result, which owes to faster on average CRe escape which creates fewer `IR-dim objects' in comparison the the \BHkappaconst model. 
    \item Surprisingly, this modeling effort reveals a remarkable \textit{insensitivity} of the global FRC to \textit{orders-of-magnitude} variable CR scattering rates (and thus emergent diffusive/streaming transport rates), though morphological and spatially resolved constraints may prove fruitful to constrain these deeply uncertain physics.
\end{itemize}

In future work, we will utilize the types of synthetic observations produced here to constrain AGN feedback and CR transport across differing galaxy properties in concert with the latest spatially-resolved observational constraints. While in this work we explored how AGN and CR physics shapes the emergence and maintenance of the \zzero\, FRC, with multi-wavelength synthetic observations in-hand, we will also explore the role observational K-corrections may play in determining the evolution of the global FRC with redshift \citep{sargent_no_2010,magnelli_far-infraredradio_2015}. 
Comparisons with these existing FRC constraints requires careful consideration of Malmquist bias effects in synthetic-observational sampling, as many of these high-z constraints contain galaxies at equivalent halo masses to those explored here \textit{at those redshifts}, rather than at $z=0$, which will require re-simulations of massive galaxies at higher redshifts with FIRE-3 physics and CR-MHD, which is a considerable simulation effort with the zoom-in approach utilized here. For instance, while some of these flux-limited observational samples appear to infer a modestly \textit{decreasing} ratio of $q \equiv $ L$_{8-1000 \mu m}$/L$_{\rm 1.4 GHz}$ with $z$, our theoretical predictions show a general trend of \qYun modestly increasing with $z$, in line with analytic predictions \citep{murphy_far-infrared-radio_2009}. 

Indeed, we see hints towards this contradictory trend with $z$ in our sample when limiting to only the brightest snapshots at a given redshift, as evinced by qualitatively weaker $z$-evolution in the rightmost panel of Figure \ref{fig:q_z_mass}, but requires larger (and more massive) simulation samples to make statistically meaningful predictions 
\citep[see also][for detailed discussion on disentangling mass and redshift trends in the FRC]{schober_model_2023}. Samples of this sort will enable comparisons to growing constraints at higher redshifts and detailed spatially-resolved comparisons at low redshifts in the ngVLA \citep{Murphy2018}, SKA \citep{SKA}, \& DSA-2000 \citep{DSA2000} era which may further shed light on the complex, non-linear, and non-thermal physics of galaxies across cosmological time.

\newpage

% as well as a motivated by `extrinsic turbulence' theory but modified ad-hoc to reproduce the correct CR spectral properties in Milky-Way Solar-Circle-like conditions. 

\section{Acknowledgements}
\begin{acknowledgments}
We wish to recognize and acknowledge the past and present Gabrielino-Tongva people and their unceded Indigenous lands upon which this research was conducted. Support for SP \&\ PFH was provided by NSF Research Grants 20009234, 2108318, NASA grant 80NSSC18K0562, and a Simons Investigator Award. Numerical calculations were run on NSF/TACC allocation AST21010, TG-AST140023, TG-PHY240164, and NASA HEC SMD-16-7592. RKC was funded by support for program \#02321, provided by NASA through a grant from the Space Telescope Science Institute, which is operated by the Association of Universities for Research in Astronomy, Inc., under NASA contract NAS 5-03127. RKC is grateful for support from the Leverhulme Trust via the Leverhulme Early Career Fellowship. ISB was supported by NASA through the Hubble Fellowship, grant HST-HF2-51525.001-A awarded by the Space Telescope Science Institute, which is operated by the Association of Universities for Research in Astronomy, Incorporated, under NASA contract NAS 5-26555. SW received support from the NASA RIA grant 80NSSC24K0838. DK was supported by NSF grant AST-2108324. The Flatiron institute is supported by the Simons Foundation.
\end{acknowledgments}


\vspace{5mm}
\facilities{TACC}


\software{astropy \citep{2013A&A...558A..33A,2018AJ....156..123A}
          }
\appendix 
\renewcommand\thefigure{A\arabic{figure}} 
\setcounter{figure}{0}

\begin{figure*}[h]
    \centering
    \includegraphics[width=1.0\textwidth,height=1.0\textheight,keepaspectratio]{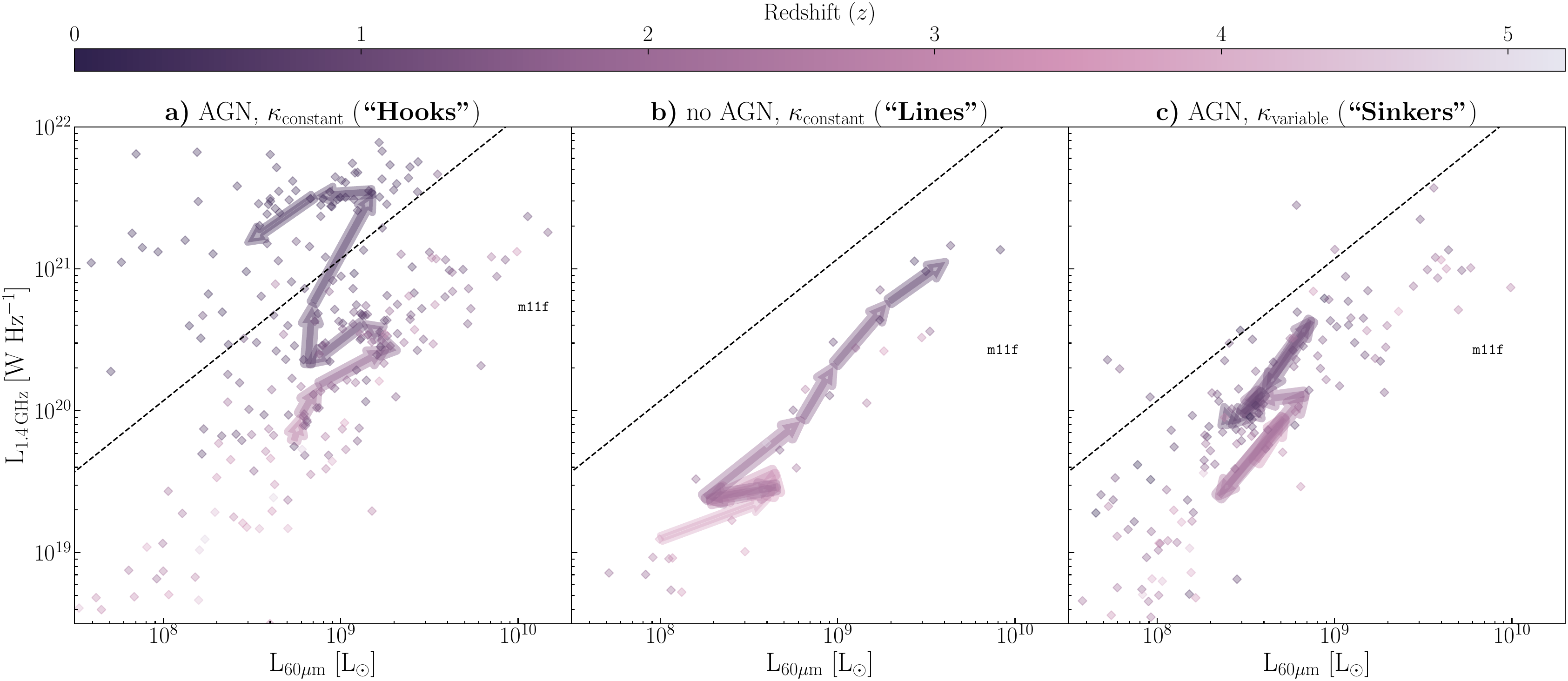}
    \caption{\textit{Evolution along FRC ``tracks,''} here shown for an individual massive dwarf, \texttt{m11f}, in the same style as Figure \ref{fig:cartoon}. The more massive dwarf galaxies in our simulation sample exhibit similar "tracks" as the \texttt{m12s} and \texttt{m13s} outlined in Figures \ref{fig:tracks} and \ref{fig:cartoon}.} 
    \label{fig:m11f}
\end{figure*}



\bibliography{full_bib}{}
\bibliographystyle{aasjournal}



\end{document}